\documentclass[a4paper,11pt]{article}
\pdfoutput=1 

\usepackage{jcappub} 

\usepackage[T1]{fontenc} 
\usepackage{ulem}
\usepackage{amsmath}

\allowdisplaybreaks[1]

\title{\boldmath Inflation with vector fields revisited: heavy entropy perturbations and primordial black holes}

\author[]{Chong-Bin Chen}

\affiliation[]{Department of Physics, Kobe University, Kobe 657-8501, Japan}

\emailAdd{chongbin@stu.kobe-u.ac.jp}

\abstract{We revisit inflation coupled with vector fields employing kinetic coupling in the comoving gauge. It is known that there is a cumulative effect $IN^2$ on the curvature power spectrum. For a large number of e-foldings $N$, this contribution is so significant that it could violate observational constraints  when the ratio of kinetic energy between the vector fields and the inflaton $I$ is not extremely small. In this paper, we explore a regime where $I\gg 1$. This regime has not been extensively explored due to the limitations of perturbative methods. We found that the entropy perturbation becomes heavy in this regime and the cumulative effect decays away on super-horizon scales. Consequently, the power spectrum retains its scale invariance in the decoupling limit. By straightforwardly integrating out the heavy modes near horizon-crossing, we derive a low-energy effective field theory describing a massless adiabatic perturbation with an imaginary speed of sound $c_s^2= -1/3$. Namely, the inflation with vector fields presents a potential mechanism for generating primordial black holes.}

\begin{document}
\maketitle
\flushbottom

\section{Introduction}
Inflationary physics has proven to be successful in phenomenologically explaining the large-scale structure of our universe. The simplest inflationary model, single-field inflation, predicts nearly scale-invariant, Gaussian, and adiabatic primordial perturbations. These predictions are confirmed by Cosmic Microwave Background (CMB) observations \cite{Smoot:1992,WMAP:2003elm,Planck:2018vyg}. However, the single-field model encounters theoretical challenges. These inflationary models are expected to be effective field theories derived from UV completed theories, such as string theory. Despite an anticipated suppression of higher-dimension irrelevant operators by UV cutoff, they play a crucial role at the low-energy scale during inflation. The UV sensitivity of inflation gives rise to issues like the $\eta$-problem \cite{Copeland:1994vg,Baumann:2014nda} and the super-Planckian problem of a single large field \cite{Lyth:1996im,Baumann:2011ws,Baumann:2014nda}, disrupting the flatness of the scalar field potential.

Study of introducing extra fields during inflation has attracted attention to evade these issues of single-field inflation.  In particular, the study of multi-scalar fields exhibiting strongly non-geodesic motion has been extensively examined in recent years \cite{Brown:2017osf,Mizuno:2017idt,Christodoulidis:2018qdw,Renaux-Petel:2015mga,Garcia-Saenz:2018ifx,Bjorkmo:2019fls,Fumagalli:2019noh}. In contrast, our focus is on U(1) gauge fields, commonly known as vector fields, as additional fields during inflation. This choice stems not only from ubiquity of gauge fields as essential components of the standard model (SM) and beyond but also from their rich phenomenological implications in cosmology \cite{Planck:2018jri,Tavecchio:2010,Neronov:2010,Tavecchio:2011,Subramanian:2015lua}. Breaking of conformal couplings between vector fields and gravity is necessary to generate sufficiently strong vector fields \cite{Subramanian:2015lua}. Analogous to the non-geodesic motion of scalar fields, vector fields should exhibit background dynamics and participate alongside the scalar fields in driving inflation. The inflationary phase undergoes a transition to new attractors if the potential is steep enough \cite{Chen:2022ccf}, which is very similar to the hyperbolic inflation \cite{Brown:2017osf}. This kind of models was firstly constructed to explain the statistical anisotropy in the CMB \cite{Watanabe:2009ct,Watanabe:2010fh,Kanno:2010nr,Yamamoto:2012tq} (or read the review \cite{Maleknejad:2012fw}). The vector field is coupled to gravity through  $f^2(\phi)F_{\mu\nu}F^{\mu\nu}$. After suitable choices of $f(\phi)$, dynamics of background vector fields provides an effective potential for the slow-roll scalar field. 

Under this new attractor, the power spectrum of the curvature perturbation is sourced by the vector fields, yielding an additional contribution 
\begin{equation}\label{Pzeta}
    \mathcal{P}_{\zeta}=\mathcal{P}_{\zeta}^{(0)}\left[1+\Theta(\theta)\mathcal{I}N_k^2\right], \ \ \ \ \ \ (\mathcal{I}\ll 1)
\end{equation}
where $N_k\equiv\ln{(-k\tau_e)}$  represents the e-folding number of observed modes elapsed
since horizon-crossing, $\tau_e$ is the end time of the inflation. $\mathcal{I}$ is a parameter that related to specific models. For one U(1) vector field, $\Theta(\theta)=24\sin^2\theta$, where $\theta$ is the angle between mode and direction of the vector field \cite{Watanabe:2010fh,Bartolo:2012sd}. The term $\mathcal{I}N_k^2$ accountable for the amplitude of statistical anisotropy and should be rigorously constrained by CMB observations to $\mathcal{I}<10^{-7}$ for a sufficiently lengthy inflationary period $N_k\sim 60$ \cite{Watanabe:2010fh,Bartolo:2012sd}. The exceedingly small parameter implies the presence of a tiny but non-zero vector field, potentially giving rise to fine-tuning challenges \cite{Naruko:2014bxa,Fujita:2017lfu}. Studies have revealed that employing multiple vector fields can mitigate the anisotropy, eventually converging to an isotropic attractor \cite{Yamamoto:2012tq}. For the isotropic configuration of a triad of vector fields, $\Theta(\theta)=16$, relaxing the constraint on $\mathcal{I}$ \cite{Gorji:2020vnh}. However, it still needs to be constrained to $\mathcal{I}<10^{-5}$ due to the tilt of the power spectrum caused by the e-folding number $N_k$. This is a cumulative effect, i.e., the curvature is sourced by the non-zero vector fields and evolves on super-horizon scales.

However, this does not imply that results for  $\mathcal{I}\sim\mathcal{O}(1)$ are undesirable. If we require vector fields to handle a sufficiently large field displacement, indicating that $\mathcal{I}$ is not a small parameter, it appears that the second term in (\ref{Pzeta}) becomes significant and violates adiabatic and scale-invariant initial conditions. However, this result lacks reliability because the perturbative method breaks down in this regime. It has been demonstrated that for $\mathcal{I}\gtrsim 0.2$, the calculation by using the in-in formalism starts to deviate from the numerical results and it is difficult to track the dynamics near and outside the horizon analytically \cite{Yamamoto:2012sq,Funakoshi:2012ym}. Numerical findings indicate that the power spectrum does not actually evolve on super-horizon scales \cite{Funakoshi:2012ym,Chen:2022ccf}. In this paper, we revisit this dilaton-gauge coupling model in the comoving gauge and thoroughly explore the significant $\mathcal{I}$ case with the assistance of the effective field theory (EFT). We will show, that the entropy perturbation is super heavy in this regime and the theory can be described by a low-energy  EFT of a massless adiabatic perturbation. $\mathcal{I}N_k^2$ does not occur in this regime; instead, it is replaced by an exponentially enhanced power spectrum, dependent solely on one parameter. This enhancement also provides a mechanism for generating primordial black holes.

\section{Inflation with vector fields}
We are considering a spatially flat Friedmann-Lemaitre-Robertson-Walker (FLRW) universe, which is characterized by the scale factor $a(t)$ and the Hubble variable is $H(t)=\dot{a}/a$, where $\dot{}\equiv d/dt$ isthe cosmological time derivative. The Arnowitt-Deser-Misner(ADM) formalism for the metric is \cite{Arnowitt:1962hi}
\begin{equation}
     ds^2=-N^2dt^2+q_{ij}\left(N^idt+dx^i\right)\left(N^jdt+dx^j\right),
\end{equation}
where $N$ is the lapse function and we choose the cosmological time $N=1$ for background, $N^i$ is the shift vector and $g_{ij}$ is the metric of the hypersurface. We consider a model of an inflaton and $U(1)$ gauge fields, which is called dilaton-gauge inflation in this paper
\begin{equation}
   S=\int d^4x\sqrt{-{g}} \left[
    -\frac{1}{2}\partial^\mu\phi \partial_\mu\phi
    -V(\phi)-\frac{1}{4}f_{ab}(\phi)F^a_{\ \mu\nu}F^{b\mu\nu}\right],
\end{equation}
where $g$ is the determinant of the metric, $V(\phi)$ is the inflationary potential and $f_{ab}$ is the kinetic coupling, which only depends on the inflaton. The strength of the vector fields is $F^a_{\ \mu\nu}=\partial_{\mu}A^a_{\ \nu}-\partial_{\nu}A^a_{\ \mu}$.

\subsection{Background}

The configuration of the gauge fields is chosen for preserving isotropy \cite{Bento:1992wy,Golovnev:2008cf}. One can consider a large number of randomly vector fields. These vector fields have about the same magnitude of order $A$ initially. Then, for example, \cite{Golovnev:2008cf}
\begin{align}
    \sum_{a}^{N\gg 1}A^{a}_{\ i}A^{a}_{\ j}\simeq \frac{N}{3}A^2\delta_{ij}+\mathcal{O}(1)\sqrt{N}A^2
\end{align}
is dominated by the $\delta_{ij}$ parts. The corrections proportional to $\sqrt{N}$ are due to stochastic random distribution of directions of the fields and they do not vanish for $i\neq j$. Moreover, if the coupling $f_{ab}$ are uniform, the isotropic configuration is an attractor in the phase space for the number
of vector fields greater than two. This has been confirmed numerically in \cite{Yamamoto:2012tq}. If the field is a SU(2) non-Abelian gauge field, it's natural to have this uniform coupling. The gauge coupling can be ignored during inflation thus SU(2) gauge field looks exactly like a triplet of U(1) gauge fields \cite{Murata:2011wv}.

Therefore in this paper we assume one dilaton field $\phi$ as the inflaton and a triplet of $U(1)$ fields and take an isotropic configuration of the background evolution, which can be realized by the following choice of $U(1)$ fields \cite{Maleknejad:2011sq,Maleknejad:2011jw}
\begin{equation}\label{iso}
    A^a_{\ 0}=0,\ \ \ \ \ A^a_{\ i}=\mathbb{A}\delta_{ai}, \ \ \ \ \ f_{ab}=f^2\delta_{ab} \ .
\end{equation}
The identification of the spatial indices and the internal indices of the field space allows us to achieve an internal gauge transformation of the three $U(1)$ gauge fields by a $O(3)$ rotation of the three spatial dimensions in real space. We note that the discussion doesn't rely on the number of the $U(1)$ fields. If we have only one $U(1)$ field and kinetic coupling with dilaton, there may be anisotropy left in our universe when the energy density of the $U(1)$ gauge field is not so small \cite{Watanabe:2010fh,Bartolo:2012sd}.

Under this configurations the Friedmann equations are given by
\begin{align}
    &M_{\text{pl}}^2H^2\left(3-\epsilon_{\phi}-\frac{3}{2}\epsilon_A\right)=V(\phi),\ \ \ \ \ \ \ \epsilon=\epsilon_{\phi}+\epsilon_A,
\end{align}
where $\epsilon\equiv-\dot{H}/H^2$, $\epsilon_{\phi}\equiv\dot{\phi}^2/(2M_{\text{pl}}^2H^2)$ and $\epsilon_A\equiv f^2\dot{\mathbb{A}}^2/(a^2M_{\text{pl}}^2H^2)$. It's also useful to define $\dot{\sigma}^2\equiv\dot{\phi}^2/2+f^2\dot{\mathbb{A}}^2/a^2$ as the ``kinetic'' energy of the matter fields. Another gravitational equation can then be written as
\begin{equation}
    H^2M_{\text{pl}}^2\epsilon=\dot{\sigma}^2.
\end{equation}
The equations of motion of the matter fields are 
\begin{align}
    &\ddot{\phi}+3H\dot{\phi}+V_{\phi}-3f_{\phi}f\frac{\dot{\mathbb{A}}^2}{a^{2}}=0,\ \ \ \ \ \ \ \ddot{\mathbb{A}}+\left(2\frac{\dot{f}}{f}+H\right)\dot{\mathbb{A}}=0, \label{eq}
\end{align}
where $V_{\phi}\equiv\partial V/\partial \phi$ and $f_{\phi}\equiv\partial f/\partial \phi$. In this paper we assume that $\dot{\phi}<0$ and $\dot{\mathbb{A}}>0$. We also define a useful variable, the ratio of the kinetic energy of $U(1)$ fields and the dilaton\footnote{The parameter $\mathcal{I}$ in \cite{Gorji:2020vnh} or in (\ref{Pzeta}) is related to $h$ in our paper as $\mathcal{I}\simeq 2h^2$ for $h\ll 1$. }
\begin{equation}
    h\equiv \sqrt{\frac{\epsilon_A}{2\epsilon_{\phi}}}.
\end{equation}
But here, the ratio $h$ needn't to be a small parameter. In fact, we will consider the larger $h\gg 1$ on small scales for PBH's production.  We will also define the slow-roll parameter $\epsilon_{h}\equiv\dot{h}/(Hh)$, $\eta_h\equiv\dot{\epsilon}_h/(H\epsilon_h)$, $\eta_\phi\equiv\ddot{\phi}/(H\dot{\phi})$ and $\eta_{A}\equiv\dot{\epsilon}_A/(H\epsilon_A)
$. 
Taking the time derivative of the Friedmann equation one obtains 
\begin{equation}
    \sqrt{2\epsilon_{\phi}}V_{\phi}=M_{\text{pl}}H^2\left(6\epsilon+2\epsilon_{\phi}\eta_{\phi}+\frac{3}{2}\epsilon_{A}\eta_{A}-3\epsilon\epsilon_A\right).
\end{equation}
Inserting $V_{\phi}$ into the equation of motion of $\phi$ we obtain
\begin{equation}\label{eq2}
    \frac{f_{\phi}}{f}M_{\text{pl}}\sqrt{2\epsilon_{\phi}}=2\left(1+\frac{1}{4}\eta_{A}-\frac{1}{2}\epsilon\right).
\end{equation}

\subsection{Quadratic action in comoving gauge}
Thanks to the isotropic background and we will work in the quadratic action and the linear fluctuations of the theory. Hence, the vector and tensor fluctuations are decoupled from the scalar ones at linear order. After using the helicity decomposition the scalar parts of the $U(1)$ fields can be written as
\begin{equation}
    A^a_{\ 0}=\partial_a\mathbb{Y},\ \ \ \ \ \ 
A^a_{\ i}=\left(\mathbb{A}+\delta\mathbb{A}\right)\delta_{ai}+\epsilon_{iab}\partial_b\mathbb{U}+\partial_i\partial_a\mathbb{M}.
\end{equation}
We do not distinguish between the indices $a$ and $i$ here because we identified the spatial rotation symmetry with the internal global $O(3)$ symmetry of the space of the gauge fields.

The scalar fluctuations of gravity and the scalar field can be written as
\begin{equation}
N=1+\alpha,\ \ \ \ N_i=\partial_i\beta,\ \ \ \ g_{ij}=a^2(t)\left(e^{2\zeta}\delta_{ij}+2\partial_i\partial_jE\right),\ \ \ \ \phi=\phi(t)+\delta\phi.
\end{equation}
We have analysed the vector and tensor perturbations in our previous work \cite{Chen:2022ccf}. The $U$ and the vector perturbations are massless (see (5.10) of \cite{Chen:2022ccf}). Therefore, there is no instability in these perturbations so we didn't discuss therm in this work. The tensor perturbations are almost the same as the single-filed inflation thus they are not interesting for our discussion  (see (5.14) of \cite{Chen:2022ccf}).

We have a total of nine scalar degrees of freedom, where four from the metric and five from the matter sector. There are four gauge freedoms of the time and spatial coordinate choice. If we have one scalar field, there are two scalar gauge freedoms in the theory. On the other hand, if a triple of gauge fields is considered, it enjoys a local $U(1)$ symmetry $A^a_{\ \mu}\rightarrow A^a_{\ \mu}+\partial_{\mu}\rho^a$, where $\rho^a$ is an arbitrary function. This allows for another one gauge freedom. Fixing each gauge freedom removes one constraint. In conclusion, we can eliminate $2\times(2+1)=6$ d.o.f, leaving $3$ remaining. 

One can fix the scalar gauge freedom of the $U(1)$ fields by $\mathbb{M}=0$ \cite{Chen:2022ccf}. The spatial flat gauge is usually used to study the dilaton-gauge theory. In this gauge, the physical d.o.f are the fluctuation of the scalar field $\delta\phi$. On the other hand, the ``comoving'' gauge is also used in the two-scalar fields \cite{Achucarro:2012sm,Garcia-Saenz:2019njm}. In this gauge, the adiabatic fluctuation is set to zero. The physical d.o.f are the curvature and entropy fluctuations. 

In this paper we adopt a similar comoving gauge for the dilaton-gauge model. The adiabatic fluctuation is given by $Q_{\sigma}\equiv-\sqrt{2}\dot{\sigma}\delta u$, where $\delta u$ is the velocity potential given by $\delta T^0_i\equiv(\rho+p)\partial_i\delta u$. For the isotropic one-form gauge fields, we have $\delta T^0_{\ i}=-\dot{\phi}\partial_i\delta\phi-(2f^2\dot{\mathbb{A}}/a^2)\partial_i\delta\mathbb{A}$  \cite{Chen:2022ccf}. Here $\rho$ and $p$ are the total energy density and pressure of the system, and in our model, we have $\rho+p=\dot{\phi}^2+2f^2\dot{\mathbb{A}}^2/a^2=2\dot{\sigma}^2$. It's useful to define a variable $\delta Q=\sqrt{2}f\delta\mathbb{A}/a$. Then the adiabatic fluctuation is given by
\begin{equation}
    Q_{\sigma}\equiv\frac{\dot{\phi}/\sqrt{2}}{\dot{\sigma}}\delta\phi+\frac{f\dot{\mathbb{A}}/a}{\dot{\sigma}}\delta Q.
\end{equation}
We choose the comoving gauge
\begin{equation}\label{comovinggauge}
    Q_{\sigma}=0,\ \ \ \ \ \ E=0.
\end{equation}
This is the unitary gauge in the gauge theory, whose Nambu-Goldstone boson is eaten by the metric. Then we have fixed all gauge freedom of the theory. The d.o.f $\mathbb{U}$ in the vector fields is the magnetic fluctuation and decouples from any other modes in our configuration. It only contributes to the isocurvature fluctuation hence we ignore it in the following discussion. We also define the entropy fluctuations. Defining \cite{Firouzjahi:2018wlp,Gorji:2020vnh}
\begin{equation}
    \cos{\mathbb{\vartheta}}\equiv\frac{\dot{\phi}/\sqrt{2}}{\dot{\sigma}},\ \ \ \ \ \ \ \ \ \ \sin{\vartheta}\equiv \frac{f\dot{\mathbb{A}}/a}{\dot{\sigma}},
\end{equation}
The adiabatic fluctuation reads $Q_{\sigma}=\cos{\vartheta}\delta\phi+\sin{\vartheta}\delta Q$. This fluctuation follows the direction of the classical ``trajectory'' of the matter fields. The entropy fluctuation is defined by
\begin{equation}
    \mathcal{F}\equiv \cos{\vartheta}\delta Q-\sin{\vartheta}\delta \phi,
\end{equation}
which is similar to the entropy fluctuation in systems of two-scalar fields. But here the entropy fluctuations is not equivalent to the isocurvature one because we still have another contribution $\mathbb{U}$, which is decoupled from the $\zeta$ and $\mathcal{F}$. 

After imposing the comoving gauge, the solutions of the equations of motion of the non-dynamical d.o.f $\alpha$, $\beta$ and $\mathbb{Y}$ can be derived (see Appendix \ref{app:action2}). Then after using the background equations (\ref{eq}) and (\ref{eq2}) to eliminate the $V$, $V_{\phi}$ and $f_{\phi}$, and inserting the constrain equations into the quadratic Lagrangian, we obtain
\begin{align}\label{L2}
    \mathcal{L}^{(2)}=&a^3 M_{\text{pl}}^2\epsilon\left(\dot{\zeta}^2-\frac{1}{a^2}(\partial \zeta)^2-m_{\sigma}^2\zeta^2\right)+\frac{a^3}{2}\left(\dot{\mathcal{F}}^2-\frac{1}{a^2}(\partial\mathcal{F})^2-m_s^2\mathcal{F}^2\right),\nonumber\\
    &-2a^3\dot{\sigma}h\left(\mathcal{A}\dot{\zeta}
    +\mathcal{B}H\zeta\right)\mathcal{F},
\end{align}
where the mass $m_{\sigma}$, $m_s$ and the coupling $\mathcal{A}$, $\mathcal{B}$ are 
\begin{align}\label{gg}
    &\ \ \ \ \mathcal{A}= 4,\ \ \ \ \ \ \ \mathcal{B}= \frac{16h^2}{1+2h^2},\ \ \ \ \ \ \ m_{\sigma}^2=\frac{16h^2}{1+2h^2}H^2,\nonumber\\
    &m_s^2= \frac{8h^4-40h^2-4}{1+2h^2}H^2-\frac{12h^4-8h^2-2}{1+2h^2}\frac{f_{\phi\phi}}{f}M_{\text{pl}}^2H^2\epsilon_{\phi},
\end{align}
where we have discarded slow-roll suppressed terms in the slow-low limit and also assumed $|V_{\phi\phi}|/H^2\ll 1$. This is the decoupling limit $\epsilon\to 0$ so that the higher-order terms of $\epsilon$ are discarded. The higher order of $\epsilon_{\phi}$ and $\epsilon_{A}$ are also vanishing because they are positive quantities as we have $\epsilon=\epsilon_{\phi}+\epsilon_{A}$. 

We find that the curvature and entropy fluctuations are both massive. The mass term of $\zeta$ is only dependent on $h$. For the mass term of $\mathcal{F}$, the second term can be regarded as the contribution from the curved field space, whose curvature is given by $R_{\text{fs}}=-f_{\phi\phi}/f$. If we assume $h=0$ initially, where we are in the single-field inflation regime, the entropy mass $m_{s}^2$ is still non-vanishing and have large negative contribution if the curvature of field space is negatively large enough. In the slow-roll regime, the second derivative of $f$ in the entropy mass becomes $f_{\phi\phi}/f\simeq f_{\phi}^2/f^2$. Then using (\ref{eq2}) the mass square of the entropy fluctuation reads
\begin{align}\label{ms}
    m_s^2\simeq -\frac{8h^2\left(3+2h^2\right)}{1+2h^2}H^2.
\end{align}
Therefore the entropy fluctuation has an imaginary mass. This is similar to the so-called geometrical destabilization \cite{Renaux-Petel:2015mga,Garcia-Saenz:2018ifx}, which is firstly studied in the multi-scalar inflation and then in the inflation with multiple vector fields \cite{Chen:2022ccf}. The differences are that the curvature fluctuation $\zeta$ is massive, and we have an additional coupling $\zeta\mathcal{F}$. Hence we next investigate the mode solutions of this Lagrangian.

\subsection{Evolution of the adiabatic mode}
We are interested in the super-horizon evolution of the adiabatic mode. We can solve the system from the equations of motion, which is shown in spatially flat gauge in \cite{Chen:2022ccf}. Here we directly find the solutions from the symmetries of Lagrangian. The quadratic action (\ref{L2}) in the long-wavelength limit, where the momentum terms are ignored, can be reduced to
\begin{equation}\label{laction}
    \mathcal{L}^{(2)}=a^3M_{\text{pl}}^2\epsilon\dot{\zeta}^2+\frac{a^3}{2}\dot{\mathcal{F}}^2-\frac{a^3}{2}m_s^2\left(\mathcal{F}+2\dot{\sigma}h\frac{\mathcal{A}\dot{\zeta}+\mathcal{B}H\zeta}{m_s^2}\right)^2,
\end{equation}
where we obtained the $\zeta\dot{\zeta}$ term through integral by parts. We see that this Lagrangian is invariant under shift transformation
\begin{align}
    \zeta\to\zeta+\zeta_0,\ \ \ \ \ \ \ \mathcal{F}\to\mathcal{F}-\frac{2\dot{\sigma}h\mathcal{B}H}{m_s^2}\zeta_0.
\end{align}
This symmetry reveals that $\zeta$ and $\mathcal{F}$ have constant solutions obeying
\begin{equation}\label{Fzeta}
    \mathcal{F}_0=-\frac{2\dot{\sigma}h\mathcal{B}H}{m_s^2}\zeta_0.   
\end{equation}
These constant solutions exist for any value of $h$. And it's obvious that these mode solutions are not the correction $IN_k^2$ in (\ref{Pzeta}). We next explore the Lagrangian for small $h$.


In the two-scalar case disscussed in \cite{Achucarro:2016fby}, where all the fields are massless after field-redefiniton hence the action has St$\ddot{\text{u}}$ckelberg-like symmetry. We find in our case, this symmetry occurs in the leading order of $h$ for small $h$. To see this, we expand the solutions of adiabatic and entropy perturbations near $h=0$
\begin{equation}
    \zeta=\zeta^{(0)}+\zeta^{(1)}+\cdots,\ \ \ \ \ \ \ \ \ \mathcal{F}=\mathcal{F}^{(0)}+\mathcal{F}^{(1)}+\cdots
\end{equation}
We see in Lagrangian (\ref{L2}), the mass terms are $\mathcal{O}(h^2)$, the $\mathcal{A}$ term is $\mathcal{O}(h)$ and the $\mathcal{B}$ term is  $\mathcal{O}(h^3)$. If we only consider super-horizon solutions up to $\mathcal{O}(h)$, the leading-order Lagrangian can be written as
\begin{align}
    \mathcal{L}^{(2)}=&\frac{a^3}{2}\left(\sqrt{2\epsilon} M_{\text{pl}}\dot{\zeta}-\sqrt{2}\mathcal{A}Hh\mathcal{F}\right)^2+\frac{a^3}{2}\dot{\mathcal{F}}^2+\mathcal{O}(h^2).
\end{align}
We found at leading order of $h$, the Lagrangian is invariant under
\begin{equation}
    \dot{\zeta}\to\dot{\zeta}+\dot{\zeta}^{(1)},\ \ \ \ \ \ \ \mathcal{F}\to\mathcal{F}+\frac{\sqrt{\epsilon}M_{\text{pl}}}{\mathcal{A}h H}\dot{\zeta}^{(1)}
\end{equation}
The first order $\dot{\zeta}^{(1)}$ is sourced from zero-th order of entropy perturbations $\mathcal{F}^{(0)}$. This source from entropy modes is a cumulative effect hence is dependent on the e-folding number after horizon-crossing
\begin{align}\label{massivemode}
    \zeta^{(1)}= \int_{N_c}^{N_e}dN\ \frac{\mathcal{A}h}{\sqrt{\epsilon}M_{\text{pl}}}\mathcal{F}^{(0)}=\frac{\mathcal{A}hN_k}{\sqrt{\epsilon}M_{\text{pl}}}\mathcal{F}^{(0)},
\end{align}
where $N_c$ is the time of horizon-crossing, $N_e$ is the end time of inflation and $N_k\equiv N_e-N_c$ is the e-folding number elapsed since horizon-crossing. For small $h$ this is a small correction to the adiabatic mode. The zero-th order entropy perturbation is decoupled with the adiabatic one and massless in the slow-roll limit. Hence the zeroth-order power spectrum of entropy perturbation is simple $\mathcal{P}_{\mathcal{F}}^{(0)}=H^2/(4\pi^2)$. Then the leading-order correction to the power spectrum of curvature perturbation can be estimated to
\begin{align}
    \delta\mathcal{P}_{\zeta}\simeq 32h^2N^2\cdot\frac{H^2}{8\pi^2\epsilon M_{\text{pl}}^2},\ \ \ \ \ (h\ll 1)
\end{align}
which is consistent with the result calculated through in-in formalism \cite{Gorji:2020vnh} and reproduces the result (\ref{Pzeta}). These correction is proportional to square of e-folding number $N_k\sim 60$, which is a large number for inflation. Hence for anisotropic configuration of vector fields, there is very strong constrain from CMB on the parameter $h$. The extreme small $h$ may lead to fine-tuning problem \cite{Naruko:2014bxa,Fujita:2017lfu}. However, these modes is heavy when $h$ is large enough hence dilute rapidly during inflation. To see this, we next explore the large $h$ case.

\section{EFT under massive entropy mode}

In this system we have two coupled d.o.f. From the quadratic Lagrangian (\ref{L2}) we can write down the linear equations of motion of the system:
\begin{align}\label{fulleom}
    \ddot{\zeta}+\left(3+\eta\right)H\dot{\zeta}+\left(\frac{k^2}{a^2}+m_{\sigma}^2\right)\zeta=&\frac{1}{a^3\epsilon}\left(a^3\frac{\dot{\sigma}h\mathcal{A}}{M_{\text{pl}}^2}\mathcal{F}\right)^{.}-\frac{\dot{\sigma}hH\mathcal{B}}{\epsilon M_{\text{pl}}^2}\mathcal{F},\nonumber\\
    \ddot{\mathcal{F}}+3H\dot{\mathcal{F}}+\left(\frac{k^2}{a^2}+m_{s}^2\right)\mathcal{F}=&-2\dot{\sigma}h\left(\mathcal{A}\dot{\zeta}+\mathcal{B}H\zeta\right).
\end{align}
\begin{figure}[tbp]
\centering
\includegraphics[scale=0.85]{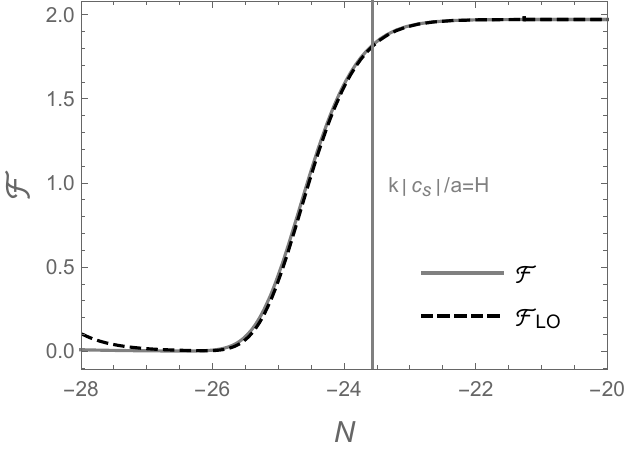}
\caption{\label{fig:EntropicF}Evolution of the entropy perturbation $\mathcal{F}$ for $h=10$. An instability before horizon-crossing and a constant evolution on super-horizon are presented. The scales are arbitrary.}
\end{figure}
In the heavy limit $|m_s^2|\gg \omega^2$ we can expand the operator as
\begin{equation}
\frac{1}{m_s^2-\square}=\frac{1}{m_s^2}+\frac{\square}{m_s^4}+\cdots,\ \ \ \ \ \  \text{where}\ \ \square\equiv -\frac{\partial^2}{\partial t^2}-3H\frac{\partial}{\partial t}+\frac{\partial^2}{a^2}.
\end{equation}
The solution of $\mathcal{F}$ is dominated by the $1/m_s^2$ term, which is given by
\begin{align}\label{Flo}
    \mathcal{F}_{\text{LO}}=-\frac{2\dot{\sigma}h(\mathcal{A}\dot{\zeta}+\mathcal{B}H\zeta)}{m_s^2}.
\end{align}
We show the consistent of the leading-order contribution of $\mathcal{F}$ in Figure \ref{fig:EntropicF} for large enough $h$. 

In the slow-roll limit, the entropy mass $m_s$ can be calculated by the (\ref{ms}). From (\ref{pUV}) the EFT is only valid under the scale
\begin{equation}
    p_{\text{UV}}=\frac{3}{4}|m_s|.
\end{equation}
Then after inserting the equation (\ref{Flo}) into the quadratic action (\ref{L2}) we obtain
\begin{equation}\label{q2action}
    \mathcal{L}^{(2)}_{\text{LO}}= a^3M_{\text{pl}}^2\frac{\epsilon}{c_s^2}\left[\dot{\zeta}^2-\frac{c_s^2}{a^2}\left(\partial \zeta\right)^2\right],
\end{equation}
where $c_s$ is the speed of sound of $\zeta$
\begin{equation}\label{cs}
    \frac{1}{c_s^2}= 1-\frac{4\left(1+2h^2\right)}{3+2h^2}.
\end{equation}
The operator from the coupling $\dot{\zeta}\mathcal{F}$ would reduce the speed of sound of the fluctuation \cite{Achucarro:2012yr}. We have also used integration by parts to eliminate the $\dot{\zeta}\zeta$ terms in the action. Then we found that the mass term of the curvature fluctuation is cancelled and $\zeta$ is conserved on super-horizon scales. We can find that in the EFT of single field, we have only this constant curvature perturbation on large scales.
\begin{figure}[tbp]
\centering
\includegraphics[scale=0.85]{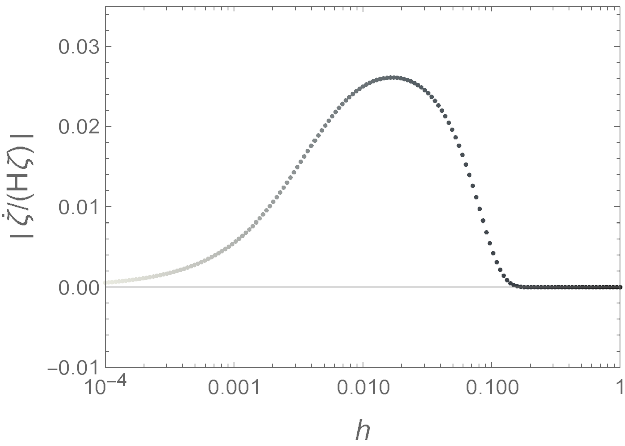}
\caption{\label{fig:Nad}The adiabaticity of the super-horizon ($|k\tau_e|=10^{-15}$) curvature perturbation. As $h$ increases from very small values, the perturbations become non-adiabatic. However, when $h$ exceeds $\mathcal{O}(0.1)$, the perturbation quickly becomes adiabatic.}
\end{figure}

The speed of sound $c_s^2\simeq -1/3$ is imaginary for a heavy entropy mass $h\gg 1$\footnote{The speed of sound is $|c_s|\sim\mathcal{O}(1)$ for large $h$. Hence the non-linear dispersion regime is narrower, and we can directly investigate the linear dispersion regime \cite{Cheung:2007st,Gwyn:2012mw,Baumann:2011su}.}. An imaginary speed of sound has been studied in two-scalar inflation recently \cite{Garcia-Saenz:2018ifx,Fumagalli:2019noh,Garcia-Saenz:2018vqf}. The entropy mass-square in two-scalar models is also imaginary, hence is tachyonic before the sound horizon crossing $p^2 c_s^2\sim H^2$. The curvature fluctuation sourced by the entropy fluctuation also experiences a transient instability before horizon crossing. After the horizon-crossing, the dynamics of fluctuations are influenced by the background. In other words, the mass term $m_{\sigma}$ and the Hubble terms become important outside the horizon.

Let's make some comments on this result:

The evolution of curvature perturbation on large scales is sourced by the non-adiabatic perturbation $\dot{\zeta}=-\delta P_{\text{nad}}H/(\rho+p)$ \cite{Garcia-Bellido:1995hsq,Wands:2000dp,Finelli:2000ya}. Then from the leading-order solution of entropy perturbation, the non-adiabatic mode can be represented as
\begin{equation}
    \delta P_{\text{nad,LO}}=\frac{m_s^2\dot{\sigma}}{Hh\mathcal{A}}\mathcal{F}_{\text{LO}}+\frac{2\dot{\sigma}^2\mathcal{B}}{\mathcal{A}}\zeta.
\end{equation}
The adiabatic initial condition ensures that the r.h.s. should be 
relatively small enough on large scales. We can see that the entropy fluctuation doesn't decay outside the horizon but is proportional to the curvature fluctuation if $\zeta=\text{constant}$ on the super-horizon. 
Unlike multi-scalar inflation, where non-decaying non-adiabatic modes will source to curvature modes so that $\zeta$ evolves on the super-horizon. In dilation-gauge inflation, we have constant $\zeta$ for non-decaying $\mathcal{F}$. In other words, For the large $h$ case, $\mathcal{F}$ doesn't play the role of non-adiabatic mode in dilaton-gauge inflation. We show this non-adiabatic occurrence for mild $h\lesssim\mathcal{O}(0.1)$ in Figure \ref{fig:Nad}. 

One of the most important findings from this result is that the cumulative modes (\ref{massivemode}) decay away for large $h$. Because for large $h$,  the modes (\ref{Flo}) on the super-horizon are exactly the constant solutions of the adiabatic mode (\ref{Fzeta}). The decaying modes can be also obtained through solving the equation of motion (\ref{fulleom}), which has been studied in the spatially flat gauge in \cite{Chen:2022ccf}. The decaying mode is $\zeta\propto (-k\tau)^{(3-\sqrt{9-96h^2})/2}$. For small $h$ we can expand this solution and find that the leading correction is proportional to e-folding number $\ln(-k\tau)=N$, which is the solution (\ref{massivemode}).

\section{Primordial black holes}
\begin{figure}[tbp]
\centering
\includegraphics[scale=0.82]{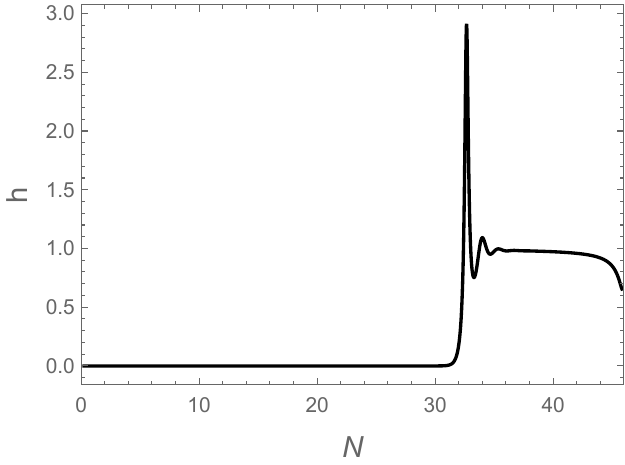}
\caption{\label{fig:example}Example of switching gauge fields on during inflation. The model with the potential $V=m^2\phi^2/2$ and coupling $f=\text{Exp}(c\phi^2/M_{\text{pl}}^2)$, where $m=10^{-6}M_{\text{pl}}$ and $c=3$.}
\end{figure}

If gauge fields are not important to the inflation, which means that inflation is driven by a single scalar field,  it may be ruinous because the higher-order loop
corrections to the scalar-field's mass state indicate that the potential has structure on scales of order of the cut-off of inflation \cite{Baumann:2014nda}. If the potential has sufficiently steep structure at some points in inflation, i.e., $V_{\phi}>3H\dot{\phi}$  \cite{Chen:2022ccf}, the gauge fields become unstable and can be switched on later. The parameter space for switching on gauge fields is wide making it not difficult for this to occur \cite{Chen:2021nkf}. We show a simple example in Figure \ref{fig:example}. If the dynamics of inflation fall into the parameter space of single-field inflation, the gauge fields will be turned off.

After switching on the gauge fields, the curvature mode experiences a exponential growth before horizon-crossing due to the imaginary speed of sound. Therefore the power spectrum is enhanced exponentially. If the vector fields switch on only at late time, the power spectrum is enhanced on small scales, making the formations of PBHs in the radiation era is possible \cite{Zeldovich:1967lct,Hawking:1971ei}. These PBHs can be candidates for dark matter \cite{Carr:1974nx,Carr:1975qj,Chapline:1975ojl} and supermassive black holes in our universe \cite{Carr:1984}. The enhancement of the power spectrum due to heavy fields is studied recently in multiple scalar-field models, and is used to discuss PBHs formation \cite{Fumagalli:2020adf,Pi:2017gih,Palma:2020ejf}. In the previous section, we have realized that it also can be achieved by using vector fields.

We approach to the PBHs without any specific models ($V(\phi)$ and $f(\phi)$). We here only consider the switch-on/off of vector fields at late time $N_f$ $only\ once$ and see the features of the power spectrum on small scales for generating PBHs. We choose the horizon-crossing time of the CMB scale $k_{\text{CMB}}=0.05\text{Mpc}^{-1}$ as $N_{\text{CMB}}=0$. Assuming that $h$ has peak at $N_f$. The curvature modes with different scales that cross the horizon at different times have different contributions to the power spectrum. We can define the corresponding scale 
\begin{equation}
    k_f|c_s|=a(N_f)H(N_f).
\end{equation}

The modes $k\gg k_f$ that cross the horizon far after the switch-off time experienced nothing. Hence the power spectrum is the same as that for single-field inflation $\mathcal{P}_{\zeta}^{(0)}=H^2/(8\pi\epsilon M_{\text{pl}}^2)$ on small scales. 

On the other hand, the modes $k\ll k_f$ that cross the horizon far before the switch-on time will be impacted by the entropy fluctuation on super-horizon scales. The entropy fluctuation doesn't decay outside the horizon and thus always transfer to the curvature fluctuation. Hence on these large scales, the power spectrum is shifted and different from that of single-field one.

The interesting scales are those where $k\sim k_f$; These modes cross the horizon near the switch-on time $N_f$ have large enhancements due to the large instability of the entropy fluctuation. We expect that the power spectrum of the curvature fluctuation has exponential enhancements near the scale $k_f$. As shown in Figure \ref{fig:EntropicF}, the mode takes a few e-folding to grow before freezing outside the horizon. Hence there are two kinds of power spectrum, the broad and the sharp of $h(N)$ \cite{Fumagalli:2020adf}. For the sharp case, the curvature modes stop growing before crossing the horizon.

\subsection{Broad case}
The EFT is valid at time $N_v$ satisfying 
\begin{equation}
    \frac{k}{a(N_v)}=\frac{3}{4}|m_s(N_v)|.
\end{equation}
Then the majority of contributions to the power spectrum come from modes with $N_v\sim N_f$. We here consider the Gaussian profile of the function 
\begin{equation}
    h_G(N)=h_0 e^{-(N-N_f)^2/(2\delta N^2)},
\end{equation}
where the profile is controlled by two parameters $(h_0,\delta N)$. The "broad" means $\delta N\gtrsim\ln{(h_0)}$, indicating that we have enough e-folding of non-zero $h(N)$ for exponential growth of the modes. On super-horizon scales, the entropy modes do not decay and still couple to the curvature ones. Hence we should discuss large-scale modes $k\ll k_f$ and small-scale modes $k\sim k_f$ separately.

\subsection*{\texorpdfstring{\textbf{Modes with} $\boldsymbol{k\gtrsim k_f}$}%
{Something with beta in it}}
To characterize the peak of the power spectrum for these modes, it's also useful to introduce a dimensionless parameter $\mathcal{I}$ \cite{Garcia-Saenz:2018vqf,Bjorkmo:2019qno}. The quantization of the curvature fluctuation with imaginary speed of sound differs significantly from that of with real one \cite{Garcia-Saenz:2018vqf}. The difference lies in the fact that, unlike the Bunch-Davies vacuum, the negative frequency modes are non-zero for initial state of the EFT due to the different commutation conditions of the mode functions. Taking the super-horizon limit, the leading contribution of the the power spectrum is
\begin{equation}\label{DeltaZeta}
    \Delta_{\zeta}(k)\equiv\frac{\mathcal{P}_{\zeta}(k)}{\mathcal{P}_{\zeta}^{(0)}(k)}\simeq e^{2\mathcal{I}(h)}\big{|}_{N_v},
\end{equation}
where the exponential factor is given by (see Appendix \ref{app:I} for derivation\footnote{Although this analytical result fixes well with the numerical one. This derivation ignores the Hubble friction in the equations of motion. After completing this paper, \cite{Christodoulidis:2023eiw} proposed a rigorous derivation when take into account the Hubble fraction terms.})
\begin{equation}
    \mathcal{I}(h)\simeq \sqrt{2}\pi\left(2-\sqrt{\frac{1+6h^2}{1+2h^2}}\right)h.
\end{equation}
For modesthat begin to experience transient instability earlier, the amplitude will be larger after crossing the horizon. And we need to use a more UV EFT to describe these modes on deeper sub-horizon scales. 
\begin{figure}[tbp]
\centering
\includegraphics[scale=0.83]{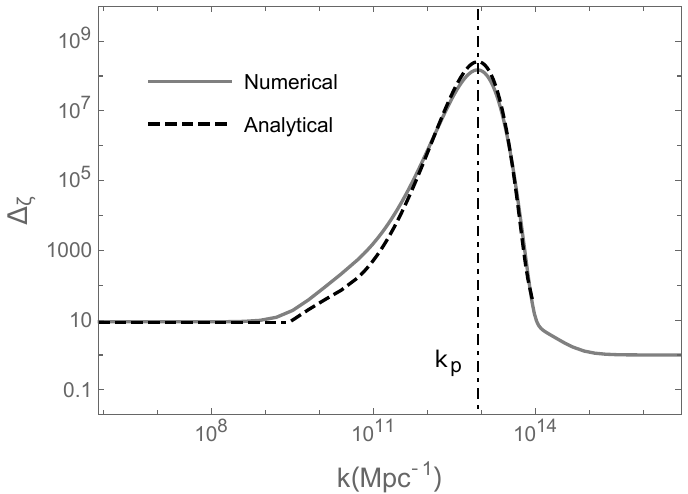}
\caption{\label{fig:PSB}The numerical and analytical results (\ref{DeltaZeta}) of the power spectrum for the Gaussian case, where $h_0=8$ and $\delta N=2$.}
\end{figure}

The scale of the peak $k_p$ is the one that has maximal $h$ at the time when the modes are growing at $N_v$, i.e., $N_{p_k}=N_v$. Thus, we have
\begin{equation}\label{kpG}
    k_p=k_f\cdot\frac{3}{4}\sqrt{\frac{8h_0^2\left(3+2h_0^2\right)}{1+2h_0^2}}.
\end{equation}
We present the numerical and analytical results of our calculation in the Figure \ref{fig:PSB}. One point to mention is that (\ref{DeltaZeta}) is not valid for all ranges of $h(N)$ because the computation is under the valid of the EFT, where the mass term of curvature is small compared to that of the heavy entropy (or in other words, for large $h(N)$). Hence, this estimation works better for $|m_s^2(N)|\gtrsim 8H^2$ in our Gaussian-profile model. 

\subsection*{\texorpdfstring{\textbf{Modes with} $\boldsymbol{k\ll k_f}$}%
{Something with beta in it}}
As Figure \ref{fig:PSB} shows, for the l.h.s of the peak, where the switch-on occurs after crossing the horizon, the power spectrum is governed by the long-range contributions. Thus we can observe a shift on large scales due to the coupling with entropy fluctuation on large scales. We have numerically found the amplitude of the power spectrum on large scale as a function of $h$ and $\delta N$
\begin{equation}\label{DeltaZetaLS}
    \Delta_{\zeta}^{\text{(LS)}}=\left(-0.00585h\cdot\delta N+0.0537h+0.333\delta N+1.95\right)^2.
\end{equation}
The amplitude of the large-scale curvature perturbation depends linearly on $h$ and $\delta N$, and the amplitude of the power spectrum $\Delta_{\zeta}^{\text{(LS)}}\gtrsim 10$, i.e., the large-scale amplitude is always greater than that of single-field inflation.

\subsection{Sharp case}
For very small $\delta N$ the Gaussian profile can be approximated by a delta function. For analytically computation, we use the top-hat profile 
\begin{equation}
    h_{\text{TH}}=h_0\left[\theta(t-t_1)-\theta(t-t_2)\right].
\end{equation}
The wave number $k_f\simeq He^{H(t_1+t_2)/2}$ is the scale at which the modes cross the horizon during the $h\neq 0$. After the switch-off, the curvature and entropy modes decouple. Then the solutions for the curvature modes are \cite{Palma:2020ejf}
\begin{align}\label{EFGH}
    \hat{\zeta}_c(t,\boldsymbol{k})=&\Big[E_{\zeta}u_k(t)+F_{\zeta}u_k^*(t)\Big]\hat{a}_{\zeta}(\boldsymbol{k})\nonumber\\
    &+\Big[G_{\zeta}u_k(t)+H_{\zeta}u_k^*(t)\Big]\hat{a}_{\mathcal{F}}(\boldsymbol{k}) +\text{h.c.}(-\boldsymbol{k}),
\end{align}
where the annihilation and creation operators of the two scalar modes satisfy the commutation relations $[\hat{a}_{\alpha}(\boldsymbol{k}),\hat{a}^{\dagger}_{\beta}(\boldsymbol{k}')]=(2\pi)^3\delta_{\alpha\beta}\delta(\boldsymbol{k}-\boldsymbol{k}')$ , $[\hat{a}^{\dagger}_{\alpha}(\boldsymbol{k}),\hat{a}^{\dagger}_{\beta}(\boldsymbol{k}')]=[\hat{a}_{\alpha}(\boldsymbol{k}),\hat{a}_{\beta}(\boldsymbol{k}')]=0$ and $u_k$ are the mode functions with Bunch-Davies initial conditions
\begin{equation}\label{BD}
    u_k(t)=\frac{iH}{\sqrt{2k^3}}\Big[1+ik\tau(t)\Big]e^{-ik\tau(t)}.
\end{equation}
The power spectrum of the curvature perturbation is then given
\begin{equation}
    \Delta_{\zeta}=|E_{\zeta}-F_{\zeta}|^2+|G_{\zeta}-H_{\zeta}|^2.
\end{equation}
The non-trivial period is when $h_{\text{TH}}\neq 0$ (which is denoted as region-II), during which the two types of modes strongly couple. We also discuss large-scale modes $k\ll k_f$ and small-scale modes $k\sim k_f$ separately.
\begin{figure}[tbp]
\centering
\includegraphics[scale=0.82]{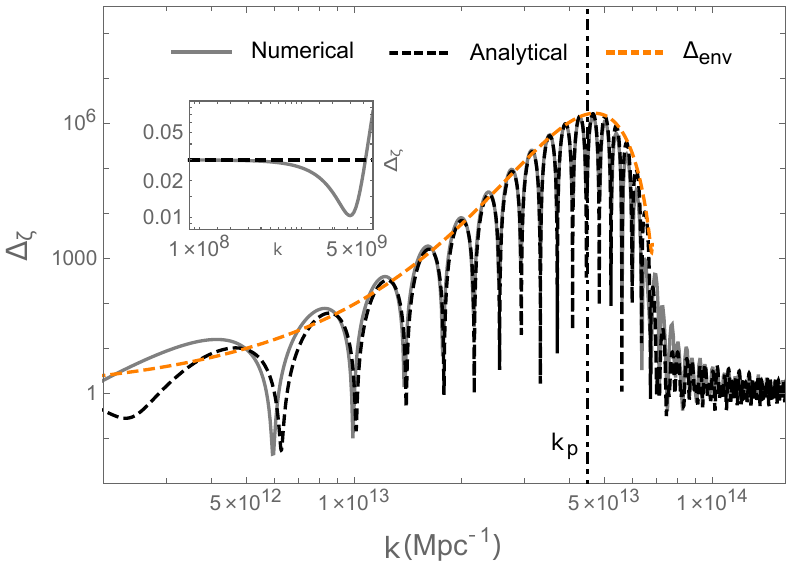}
\caption{\label{fig:PSS}The numerical and analytical results (\ref{EFGHfull}) of the power spectrum for the top-hat case, where $\delta N=0.4$ and $h_0=24$. The envelope (\ref{DeltaZetaenv}) is shown as orange dashed curve.}
\end{figure}

\subsection*{\texorpdfstring{\textbf{Modes with} $\boldsymbol{k\gtrsim k_f}$}%
{Something with beta in it}}
We use the conformal time as a function of $t$: $\tau(t)=-e^{-Ht}/H$. During $t_1$ and $t_2$ with $\delta t\equiv t_2-t_1\ll H^{-1}$, one can ignore the Hubble friction in the equation of motions. After $t>t_1$, the coupling is switch-on. The duration of $h\neq 0$ is too short, $\delta t\ll H^{-1}$, that we can ignore the evolution of the scale factor. Hence we can ignore the Hubble frictions and curvature mass. We then use the equations of motion (\ref{seom}) and the WKB-type solutions:
\begin{align}
    \hat{\zeta}_c^{\text{II}}(t,\boldsymbol{k})=\sum_{\pm}&\Big(A_{\pm}^{\zeta}e^{i\omega_{\pm}t}+B_{\pm}^{\zeta}e^{-i\omega_{\pm}t}\Big)\hat{a}_{\zeta}(\boldsymbol{k})+\Big(C_{\pm}^{\zeta}e^{i\omega_{\pm}t}+D_{\pm}^{\zeta}e^{-i\omega_{\pm}t}\Big)\hat{a}_{\mathcal{F}}(\boldsymbol{k})\nonumber\\
    &+\text{h.c.}(-\boldsymbol{k}),\nonumber\\
    \hat{\mathcal{F}}^{\text{II}}(t,\boldsymbol{k})=\sum_{\pm}&\Big(A_{\pm}^{\mathcal{F}}e^{i\omega_{\pm}t}+B_{\pm}^{\mathcal{F}}e^{-i\omega_{\pm}t}\Big)\hat{a}_{\zeta}(\boldsymbol{k})+\Big(C_{\pm}^{\mathcal{F}}e^{i\omega_{\pm}t}+D_{\pm}^{\mathcal{F}}e^{-i\omega_{\pm}t}\Big)\hat{a}_{\mathcal{F}}(\boldsymbol{k})\nonumber\\
    &+\text{h.c.}(-\boldsymbol{k}),
\end{align}
where the frequencies are given by (\ref{omega}). From the equations of motion, we have relations for the coefficients
\begin{gather*}
    A_{\pm}^{\mathcal{F}}(C_{\pm}^{\mathcal{F}})=i\frac{\omega_{\pm}^2-p^2}{\sqrt{2}h\mathcal{A}H\omega_{\pm}}A_{\pm}^{\zeta}(C_{\pm}^{\zeta}),\nonumber\\
    B_{\pm}^{\mathcal{F}}(D_{\pm}^{\mathcal{F}})=-i\frac{\omega_{\pm}^2-p^2}{\sqrt{2}h\mathcal{A}H\omega_{\pm}}B_{\pm}^{\zeta}(D_{\pm}^{\zeta}).
\end{gather*}

During the short region-II, where the scale factor is nearly invariant, we have $a=a(N_f)=k_f/H$, hence $p=(k/k_f)H$ in the above equations. There are eight unknown coefficients $A_{\pm}^{\zeta}$, $B_{\pm}^{\zeta}$, $C_{\pm}^{\zeta}$ and $D_{\pm}^{\zeta}$ in the region-II, and eight boundary conditions at $t_1$, so these coefficients can be completely solved. Then at $t_2$, we also have eight boundary conditions and the coefficients of (\ref{EFGH}) can be solved to (see Appendix \ref{app:Bogoliubov})
\begin{align}
    E_{\zeta}&\simeq \frac{\left(S_-+\kappa\right)^2}{4\kappa S_-}e^{i2\kappa\sinh(\delta N/2)-iS_-\delta N},\nonumber\\
    F_{\zeta}&\simeq-\frac{S_-^2-\left(\kappa+i\right)^2}{4k}e^{i2\kappa\cosh(\delta N/2)}\cdot i2\sin\left(S_-\delta N\right),\nonumber\\
    G_{\zeta}&\simeq \sqrt{2}h\frac{\left(\kappa^2+iS_-+\kappa S_-\right)\left(i-\kappa-S_-\right)}{\kappa S_-S_+^2}e^{i2\kappa\sinh(\delta N/2)-iS_-\delta N},\nonumber\\
    H_{\zeta}&\simeq \sqrt{2}h\frac{\left(\kappa^2+iS_-+\kappa S_-\right)\left(i+\kappa-S_-\right)}{\kappa S_-S_+^2}e^{i2\kappa\cosh(\delta N/2)-iS_-\delta N},
\end{align}
where $\kappa\equiv k/k_f$, $\delta N\equiv H(t_2-t_1)$ and $S_{\pm}\equiv \omega_{\pm}/H$. We show the agreement between numerical and completed analytical results (see appendix \ref{app:Bogoliubov}) in Figure \ref{fig:PSS}. The power spectrum has exponential enhancement near $k_f$ and also rapidly oscillate at these scales. We found that $E_{\zeta}$ and $F_{\zeta}$ are dominated, so the power spectrum can be simply written as $\Delta_{\zeta}\simeq\Delta_{\text{env}}\times\left(\text{rapid oscillation}\right)$, where the envelope is
\begin{equation}\label{DeltaZetaenv}
    \Delta_{\text{env}}=\left|\frac{e^{-i2S_{-}\delta N}\left(\kappa^2-|S_-|^2\right)}{4|S_{-}|^2}\right|
\end{equation}
and the rapid oscillation is characterized by $\text{Exp}\left(i2\kappa e^{\pm\delta N/2}\right)$, which is only related to the period of non-zero $h_{\text{tp}}$. The enhancement of the power spectrum is provided by the low-frequency $\omega_-$(see Appendix \ref{app:EFT}) and period $\delta N$ together. The scale of the peak can be nearly determined by the exponential factor in $\Delta_{\text{env}}$, which is estimated by
\begin{equation}\label{kpTH}
    k_p=k_f\cdot\frac{h_0\sqrt{30+104h_0^2+56h_0^4}}{2(1+2h_0^2)}.
\end{equation}
These calculations are only valid near the scale $k\gtrsim k_f$ and for $k\ll k_f$ $\Delta_{\zeta}=1$, which miss the matching with the numerical results.

\subsection*{\texorpdfstring{\textbf{Modes with} $\boldsymbol{k\ll k_f}$}%
{Something with beta in it}}
\begin{figure}[tbp]
\centering
\includegraphics[scale=0.8]{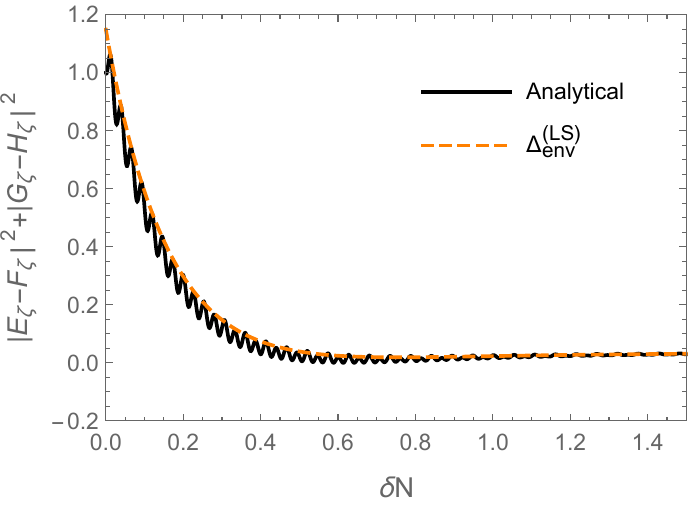}
\caption{\label{fig:PSSLdeltaN}In the top-hat case, the analytical result of the large-scale power spectrum as a function of $\delta N$ (\ref{EFGHls}), where $h_0=24$. The envelope (\ref{DeltaZetaenvLS}) is shown as the orange dashed curve.}
\end{figure}
On large scales, the momentum terms in the equations of motion can be ignored, and the mass terms and all coupling terms become important. The equations of motion can be solved as 
\begin{align}
    \hat{\zeta}_c^{\text{II}}(\tau,\boldsymbol{k})=&\sum_{i=\zeta,\mathcal{F}}\left[c_0^{i}+c_1^{i}x^{3}+c_{+}^{i}x^{1+p_{+}}+c_{-}^{i}x^{1+p_{-}}\right]\hat{a}_{i}(\boldsymbol{k})+\text{h.c.}(-\boldsymbol{k}),\nonumber\\
    \hat{\mathcal{F}}^{\text{II}}(\tau,\boldsymbol{k})=&\sum_{i=\zeta,\mathcal{F}}\left[d_0^{i}+d_1^{i}x^{3}+d_{+}^{i}x^{1+p_{+}}+d_{-}^{i}x^{1+p_{-}}\right]\hat{a}_{i}(\boldsymbol{k})+\text{h.c.}(-\boldsymbol{k}),
\end{align}
where $x\equiv-k\tau$ and the coefficients satisfy
\begin{align}
    d_0^i=\frac{2\sqrt{2}h}{3+2h^2}c_0^i,&\ \ \ \ \ d_1^i=-\frac{1}{\sqrt{2}h}c_1^i,\nonumber\\
    d_{+}^i=\frac{p_{-}(1+2h^2)-2}{2\sqrt{2}h(3+4h^2)}c_{+}^i,&\ \ \ \ \ d_{-}^i=\frac{p_{+}(1+2h^2)-2}{2\sqrt{2}h(3+4h^2)}c_{-}^i.
\end{align}
We have eight unknown coefficients $c_{0}^{i}$, $c_{1}^{i}$, $c_{+}^{i}$ and $c_{-}^{i}$ ($i=\zeta,\mathcal{F}$) in region-II and eight boundary conditions at $t_1$, so all of these coefficients can be completely solved. Then at $t_2$ the coefficients of (\ref{EFGH}) can also be solved\footnote{We found that this computation breaks when $x\lesssim 10^{-5}$.}. The results are extremely tedious, but we can simplify them under some approximations, see appendix \ref{app:Bogoliubov}. The results should be real because the perturbations become classical on the super-horizon. So we can only consider the real parts
\begin{align}\label{EFGHls}
    \text{Re}[E_{\zeta}]&\simeq \frac{1}{27}\left[-5+ 32\cosh{\left(3\delta N\right)}\right],\nonumber\\
    \text{Re}[F_{\zeta}]&\simeq \frac{32}{27}\sinh\left(3\delta N\right),\nonumber\\
    \text{Re}[G_{\zeta}]&\simeq \frac{e^{-3\delta N/2}}{3\sqrt{3}}\left(2e^{3\delta N}+1\right)\sin{\left(2\sqrt{6}h \delta N\right)},\nonumber\\
    \text{Re}[H_{\zeta}]&\simeq \frac{e^{-3\delta N/2}}{3\sqrt{3}}\left(2e^{3\delta N}-1\right)\sin{\left(2\sqrt{6}h \delta N\right)}.
\end{align}

We found that, for fixed $\delta N$, $|G_{\zeta}-H_{\zeta}|$ oscillates around zero when changing $h$. The oscillation is characterized by $\sin(2\sqrt{6}h\delta N)$ and is not significant in amplitude. This is because, on large scales, the dynamics of the curvature perturbation is almost constant and weakly depends on the frequency. For large $h$, the dynamics is determined by an EFT of massless curvature perturbation with sound speed $c_s\simeq-1/3$. On the other hand, the amplitudes of $|E_{\zeta}-F_{\zeta}|$ and $|G_{\zeta}-H_{\zeta}|$ exponentially depend on $\delta N$. We can smooth out the oscillation and only estimate the envelope of the power spectrum as
\begin{equation}\label{DeltaZetaenvLS}
    \Delta_{\text{env}}^{\text{(LS)}}= \frac{1}{729}\left(25+1024e^{-6\delta N}-212e^{-3\delta N}\right),
\end{equation}
as Figure \ref{fig:PSSLdeltaN} shows. We can see $\Delta_{\zeta}<1$, in other words, the amplitude of the power spectrum on large scales is always smaller than that of single-field inflation and is not sensitive to $h$ for sharp case. We should mention that this result is also valid for broad case with a top-hat profile of $h$. For $\delta N\gg1$, we have $ \Delta_{\text{env}}^{\text{(LS)}}\simeq0.034$, which is not sensitive to $\delta N$ either.

\subsection{PBH formation}

After treating the power spectrum on large and small scales for the broad and sharp case, we can also compare them with the amplitude of the curvature power spectrum required for PBH formation. The fraction of PBHs at the formation time can be estimated in a model-independent way \cite{Carr:2009jm}
\begin{equation}\label{betapbh}
    \beta_{\text{PBH}}=3.7\times 10^{-9}\left(\frac{\gamma}{0.2}\right)^{-1/2}\left(\frac{g_{*,f}}{10.75}\right)^{1/4}\left(\frac{M_{\text{PBH}}}{M_{\odot}}\right)^{1/2}f_{\text{PBH}},
\end{equation}
where $\gamma$ can be calculated analytically as $\gamma=0.2$ \cite{Carr:1975qj}, $g_{*,f}$ is the effective number of relativistic degrees of freedom at formation time, $M_{\text{PBH}}$ is the mass of PBHs at formation time, and $f_{\text{PBH}}\equiv\Omega_{\text{PBH}}/\Omega_{\text{DM}}$ is the present-day fraction of PBH density.

On the other hand, the $\beta$ can be related to the initial conditions of the matter power spectrum through the collapse theory of Newtonian gravity. Acommonly used model is the Press-Schechter one \cite{Press:1973iz}. After assuming a Gaussian distribution of overdensity and taking the critical density contrast as the threshold $\delta_c=0.4$ \cite{Harada:2013epa}, $\beta$ can be obtained as
\begin{equation}\label{betapbhPS}
    \beta_{\text{PBH}}\simeq \frac{\sigma}{\sqrt{2\pi}\delta_c}\text{Exp}{\left(-\frac{\delta_c^2}{2\sigma^2}\right)},
\end{equation}
where $\sigma^2\ll\delta_c^2$ is the variance of the distribution. A smaller $\sigma$ implies fewer contrasts exceeding the critical value, resulting in a smaller $\beta_{\text{PBH}}$. The variance can be calculated from the primordial power spectrum as \cite{Young:2014ana}
\begin{equation}
    \sigma^2(R)\simeq \frac{16}{81}\int d\ln q\ W^2(q,R)(qR)^4\mathcal{P}_{\zeta}(q),
\end{equation}
where the window function $W(q,R)$ is usually selected as Gaussian or top-hat, and the comoving horizon size of the PBH formed is related to the mass as $R^{-1}= k_p$, where \cite{Sasaki:2018dmp}
\begin{equation}
    M_{\text{PBH}}= 30M_{\odot}\left(\frac{\gamma}{0.2}\right)\left(\frac{g_{*,f}}{10.75}\right)^{-1/6}\left(\frac{k_p}{2.9\times 10^{5}\text{Mpc}^{-1}}\right)^{-2}.
\end{equation}
For inflation with a triad of vector fields, where both large-scale and PBH-scale are not trivial, the enhancement at the PBH scale comparing to the CMB scale is
\begin{equation}
    \Delta\mathcal{P}_{\zeta}\equiv\frac{\mathcal{P}_{\zeta}(k_p)}{\mathcal{P}_{\zeta}(k_{\text{CMB}})}=\frac{\Delta_{\zeta}(k_p)}{\Delta_{\zeta}^{\text{(LS)}}}.
\end{equation}
The power spectrum at the CMB scale is $\mathcal{P}_{\zeta}(k_{\text{CMB}})=2\times 10^{-9}$. For the broad case, this is given by (\ref{DeltaZeta}) and (\ref{DeltaZetaLS}), while for the sharp case, the rapid oscillation will be smoothed out by the window function and we can use (\ref{DeltaZetaenv}) and (\ref{DeltaZetaenvLS}).

\begin{figure}[tbp]
\centering
\includegraphics[scale=0.55]{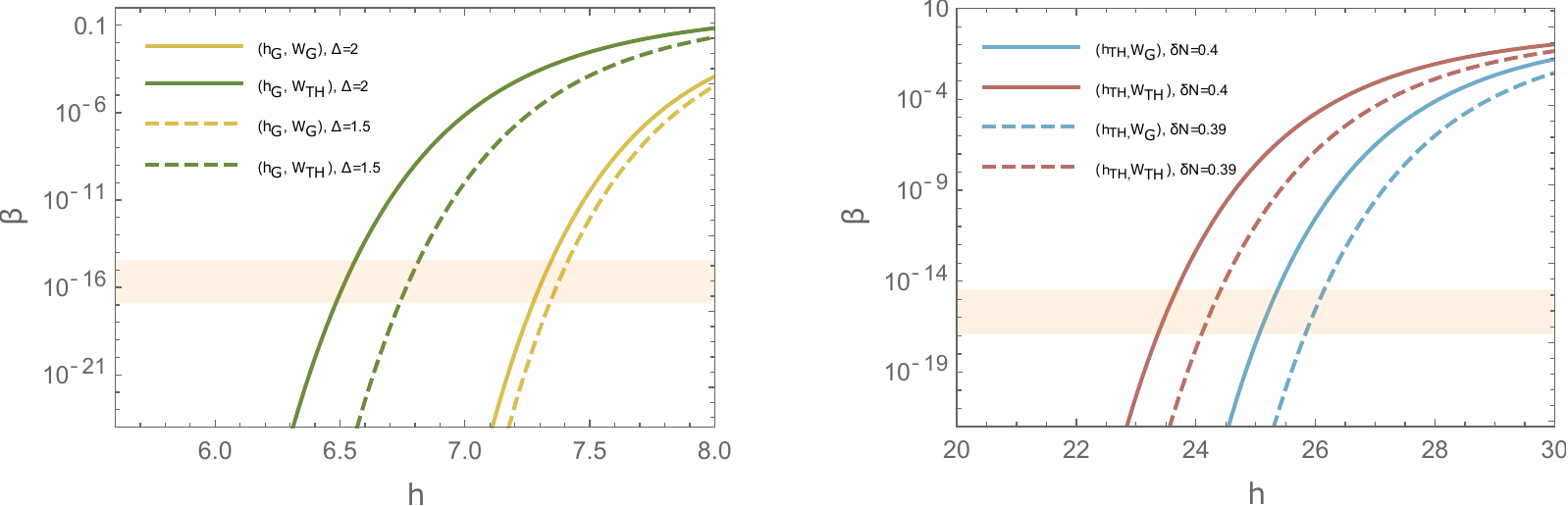}
\caption{\label{fig:Betatot}The fraction of PBHs at formation $\beta_{\text{PBH}}$ v.s. the height of the peaks $h$. The solid curves denote the cases of $\delta N=2$ for $h_{\text{G}}$ and $\delta N=0.4$ for $h_{\text{TH}}$, while the dashed curves denote the cases of smaller $\delta N$. THe orange region denotes the regime where $f_\text{PBH}=1$ between $10^{-17}M_{\odot}<M_{\text{PBH}}<10^{-12}M_{\odot}$.}
\end{figure}
If we consider the window where the PBHs can be the a candidate of dark matter: $10^{-17}M_{\odot}<M_{\text{PBH}}<10^{-12}M_{\odot}$ and $f_{\text{PBH}}=1$, from (\ref{betapbh}) we obtain $1.2\times10^{-17}<\beta_{\text{PBH}}<3.7\times10^{-15}$. Then, from (\ref{betapbhPS}) the variance for PBH formation in this mass regimes is $0.047<\sigma<0.051$. When the power spectrum is peaked at $k_p$, the variance is reduced to $\sigma^2\simeq \mathcal{P}_{\zeta}(k_p)/5$. It seems that we need $\mathcal{P}_{\zeta}(k_p)\sim 10^{-2}$ between $10^{12}\text{Mpc}^{-1}<k_p<10^{15}\text{Mpc}^{-1}$ to generate sufficient PBHs. However, because the profile of the peak is very sensitive to $h$ and $\delta N$, and the large-scale power spectrum also depends on these parameters. The fraction $\beta_{\text{PBH}}$ is exponentially dependent on the power spectrum. Slightly changing the profile of the peak will lead to a significant change in $\beta_{\text{PBH}}$. We show how $\beta_{\text{PBH}}$ depends on parameters in Figure \ref{fig:Betatot} for broad Gaussian peaks and  sharp top-hat peaks. We have used $R=k_p^{-1}$, which are the scales of the peaks and can be separately obtained through (\ref{kpG}) and (\ref{kpTH}) for Gaussian and top-hat examples. The $\beta_{\text{PBH}}$ decays rapidly for smaller parameters, i.e., smaller height and fatness of the peaks. Also, the difference in using different window functions is also distinguishable.

\begin{figure}[tbp]
\centering
\includegraphics[scale=0.9]{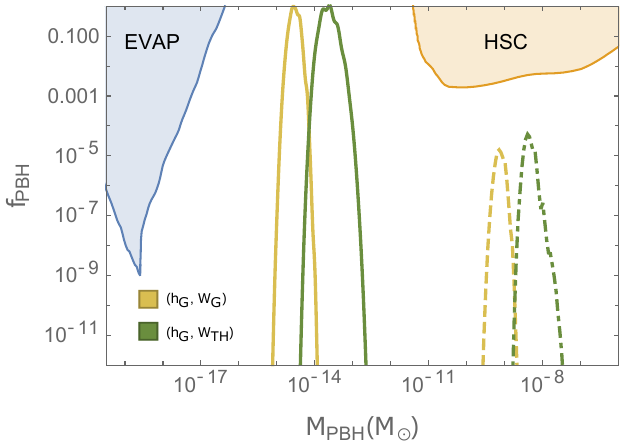}
\caption{\label{fig:fG}Abundance of PBHs as dark matter for broad($\delta N\gtrsim 1$) Gaussian profile of peak $h_\text{G}$.}
\end{figure}
\begin{figure}[t]
\centering
\includegraphics[scale=0.9]{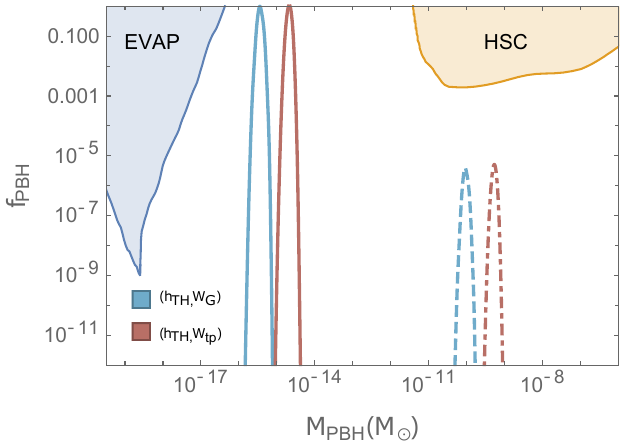}
\caption{\label{fig:fTH}Abundance of PBHs as dark matter for sharp($\delta N\lesssim 1$) top-hat profile of peak $h_\text{TH}$.}
\end{figure}
The mass spectrum of PBHs are shown in Figure \ref{fig:fG} for broad Gaussian peaks and in Figure \ref{fig:fTH} for sharp top-hat peaks. We choose the parameters of these examples as following table.
\begin{table}[ht]
\begin{center}
 \setlength{\tabcolsep}{4.8mm}{
\begin{tabular}{cc|cccc}
\hline
                                             &            & ($h_{\text{G}}$, $W_{\text{G}}$) & \multicolumn{1}{l}{($h_{\text{G}}$, $W_{\text{TH}}$)} & \multicolumn{1}{l}{($h_{\text{TH}}$, $W_{\text{G}}$)} & \multicolumn{1}{l}{($h_{\text{TH}}$, $W_{\text{TH}}$)} \\ \hline
\multicolumn{1}{c|}{{Solid}}  & $h$        & 7.258                            & 5.997                                                 & 24.546                                                & 19.285                                                 \\
\multicolumn{1}{c|}{}                        & $\delta N$ & 2                                & 2                                                     & 0.4                                                   & 0.4                                                    \\ \hline
\multicolumn{1}{c|}{{Dashed}} & $h$        & 7.200                            & 5.997                                                 & 24.246                                                & 19.285                                                 \\
\multicolumn{1}{c|}{}                        & $\delta N$ & 2                                & 1.5                                                   & 0.4                                                   & 0.395                                                  \\ \hline
\end{tabular}}
\end{center}
\end{table}
For the solid curves, we choose the parameters so that the PBHs are responsible for the total dark matter $f_{\text{PBH}}=1$ in our universe. The time of non-vanishing vector fields are chosen as $N_f=31.5$ for all solid examples to avoid constraints from observations of the evaporation of PBHs due to Hawking radiation (EVAP) \cite{Carr:2009jm} and microlensing of stars in M31 by Subaru Hyper SuprimeCam (HSC) \cite{Niikura:2017zjd}. For the dashed curves, we choose different parameters so that $f_{\text{PBH}}<1$ and also different $N_f=25.3$, which corresponds to the formation of larger PBH mass. We have seen in Figure $\ref{fig:Betatot}$ the abundance is sensitive to these parameter. As the first and the third column show, a mere 1\% decrease in $h$ leads to a drastic drop in abundance from 1 to $10^{-5}$. For the top-hat profile of the peak $h_{\text{TH}}$, the abundance is also sensitive to $\delta N$ (\ref{DeltaZetaenv}). So we can see in the last column of table, 1\% decrease in $\delta N$ also reduces $f_{\text{PBH}}$ to $10^{-5}$.

We also compare the results of using Gaussian and top-hat window functions. The choice of the method of smoothing out the fluctuations increases the uncertainty in the results of PBH formation \cite{Ando:2018qdb,Young:2019osy}. We demonstrate a significant difference in the parameters required when using different window functions. For example, in the first row of the table, to generate the same abundance of PBHs, the required $h$ differs by up to 20\% when using Gaussian and top-hat window functions. We usually need a larger primordial density in Gaussian smoothing. Furthermore, we also find that using different window functions results in variations in the peak position of $f_{\text{PBH}}$. The mass of formation is usually greater when using the top-hat window function. 
In these examples, this difference reaches up to $M_{\text{peak,TH}}/M_{\text{peak,G}}\sim 10$.

\section{Discussion}
In this paper we explore the inflation kinetically coupled to vector fields through function $f(\phi)$ in the large $h$ regime. We revisit this model in the comoving gauge, encompassing coupled curvature and entropy perturbations. By directly analyzing the symmetries of the Lagrangian, we identify solutions for these perturbations on super-horizon scales. We find that, in addition to the solution contributing to the $IN^2$ correction due to the St$\ddot{\text{u}}$ckelberg-like symmetry when $h$ is very small, there exists constant solutions (\ref{Fzeta}) for the system at any value of $h$. However, when $h$ is large, the entropy perturbation has significant and imaginary mass (\ref{ms}). In this regime, the leading-order solution of the entropy perturbation on super-horizon scales is given by the constant solution (\ref{Fzeta}). This implies that the solution contributing to the $IN^2$ correction quickly decays on super-horizon scales, leaving only the constant solution. This maintains a scale-invariant power spectrum. In the effective field theory (\ref{q2action}),  the curvature perturbation is massless and has an imaginary speed of sound, leading to an exponential enhancement before crossing the horizon. This is similar to the case of hyperbolic inflation, but the difference lies in the fact that outside the horizon, entropy perturbations do not decay and remain strongly coupled to curvature perturbations, causing in the large-scale modes to depend on the dynamics of the vector fields.

In addition to driving inflation, the enhancement of the power spectrum can also serve as a mechanism for the generation of primordial black holes (PBHs). This requires a strong excitation of the vector field during the late stages of inflation. We discussed two toy examples of $h$: Gaussian and top-hat. The Gaussian is employed to discuss the broad case, while the top-hat is used to address the sharp case, as both yield more precise analytical results. Due to the non-decaying entropy perturbations, the large-scale power spectrum is shifted compared to the single-field case. The amplitude of this shift depends on the dynamics of the vector fields, i.e., $\delta N$ and $h$, as shown in (\ref{DeltaZetaLS}) and (\ref{DeltaZetaenvLS}). For the sharp top-hat case, this amplitude is only a function of $\delta N$, allowing for an analytical calculation of the shape of the power spectrum. Given the exponential dependence of the power spectrum's shape on these parameters and the subsequent exponential dependence of PBH abundance on the power spectrum's shape, the PBH abundance is highly sensitive to these parameters. Whether it's $h$ or $\delta N$, even slight variations can significantly reduce the abundance of PBHs. Certainly, this sensitivity is also related to the specific details of PBH collapse. For instance, we demonstrated that, under the same $\delta_c$, smoothing the same fluctuations in different ways can lead to vastly different results, not only in PBH abundance but also in the masses of the PBH formation. For example, compared to a Gaussian window function, using a top-hat window function typically results in PBHs with smaller masses. Further investigation is required to address this question comprehensively.

Now, we delve into another regime, characterized by a relatively large energy of the vector field. In this scenario, the contributions to inflationary predictions differ significantly from the case where $h$ is very small. Many questions need to be reconsidered in this regime, such as the anisotropy for a single vector field, and non-Gaussianity. We posit that the discussions in this paper are also applicable to a single vector field. In this case, the $IN^2$ term in the power spectrum, which served as a contribution to anisotropy in the small $h$ case, is replaced by other constant modes.  It is imperative for us to calculate the anisotropy in this regime. Additionally, when considering a charged field as the inflaton, whether the perturbative Schwinger effect from the electric field is crucial to inflationary predictions is also a question that warrants consideration. Regarding PBHs, we focus on the isotropic configuration of the vector fields as an example. Although it bears similarities to hyperbolic inflation, in general, vector fields themselves exhibit phenomenology distinct from scalar fields, such as anisotropy for a single vector field and chirality for symmetry breaking. These differences manifest in the generated PBHs and the induced gravitational waves. Recently observations from 15-year NANOGrav
data set reveals information about low-frequency stochastic gravitational wave background \cite{NANOGrav:2023gor}. PBHs can be one of the sources of this background \cite{NANOGrav:2023hvm}. The formation of PBHs in single-field up to one-loop correction is still controversial \cite{Kristiano:2022maq,Riotto:2023hoz,Kristiano:2023scm,Riotto:2023gpm,Firouzjahi:2023ahg,Choudhury:2023vuj,Choudhury:2023jlt,Choudhury:2023rks}. The single-field EFT with a imaginary $c_s$ should also be discussed. We intend to present these studies in the near future.

\acknowledgments

I would like to thank Jiro Soda for helpful discussions. C-B.\,C. was supported by Japanese Government (MEXT) Scholarship and China Scholarship Council (CSC).

\appendix
\section{Derivation of the quadratic action}
\label{app:action2}
In this appendix we show the full quadratic action of the inflation with a triad of vector fields. We will fix the gauge freedoms of gravity as comoving gauge and then eliminate the non-dynamical degrees of freedom. 

We can write down the Lagrangian of gravity, scalar field and $U(1)$ fields as following
\begin{gather*}
    \mathcal{L}_{g}=N\sqrt{q}\left[\frac{M_{\text{pl}}^2}{2}\left(^{(3)}R+K_{ij}K^{ij}-K^2\right)\right],\\
    \mathcal{L}_{\phi}=\sqrt{q}\left[\frac{1}{2N}\pi^2-\frac{N}{2}\partial_i\phi\partial^i\phi-NV(\phi)\right],\\
    \mathcal{L}_{A}=\sqrt{q}\left[\frac{f_{ab}}{2N}q^{ik}(E^a_{\ i}+F^a_{\ ij}N^j)(E^b_{\ k}+F^b_{\ kl}N^l)-\frac{Nf_{ab}}{4}q^{ik}q^{jl}F^a_{ij}F^b_{kl}\right],
\end{gather*}
where we have defined $\pi=\dot{\phi}-N^j\phi_{|j}$ and $q_{ij}$ is the induced metric of space-like hypersurface and $E^a_i\equiv F^a_{0i}$. The extrinsic curvature and the spatial curvature are
\begin{gather*}
    K_{ij}=\frac{1}{2N}\left(\dot{q}_{ij}-2N_{(i|j)}\right),\\
    {^{(3)}R}=\left(q_{ij,kl}+q_{mn}{^{(3)}\Gamma^m_{ij}} {^{(3)}\Gamma^n_{kl}}\right)\left(q^{ik}q^{jl}-q^{ij}q^{kl}\right)
\end{gather*}

The quadratic action is given by
\begin{align}\label{q1}
    \mathcal{L}^{(2)}=&\frac{a^3}{2}\Bigg\{\Big[-6M_{\text{pl}}^2H^2\alpha+\dot{\phi}^2\alpha-2V_{\phi}\delta\phi-2\dot{\phi}\delta\dot{\phi}+3f^2\frac{\dot{\mathbb{A}}^2}{a^2}\alpha-6ff_{\phi}\frac{\dot{\mathbb{A}}^2}{a^2}\delta\phi\nonumber\\
    &-\sqrt{2}f\frac{\dot{\mathbb{A}}}{a}\Big(3\delta\dot{Q}+3(H-\frac{\dot{f}}{f})\delta Q\Big)+2f^2\frac{\dot{\mathbb{A}}}{a^2}\partial^2\mathbb{Y} \Big]\alpha+\frac{2}{a^2}\Big(\dot{\phi}\delta\phi-2M_{\text{pl}}^2HA\nonumber\\
    &+\sqrt{2}f\frac{\dot{\mathbb{A}}}{a}\delta Q\Big)\partial^2\beta+(\delta\dot{\phi})^2-\frac{1}{a^2}(\partial_i\delta\phi)^2-V_{\phi\phi}\left(\delta\phi\right)^2+\frac{3}{2}\Big[(\delta\dot{Q})^2\nonumber\\
    &+2(H-\frac{\dot{f}}{f})\delta Q\delta\dot{ Q}+(H-\frac{\dot{f}}{f})^2(\delta Q)^2\Big]-\frac{1}{a^2}(\partial_i \delta Q)^2+\frac{f^2}{a^2}\partial^2\mathbb{Y}\partial^2\mathbb{Y}\nonumber\\
    &-\sqrt{2}\frac{f}{a}\Big(\delta\dot{Q}+(H-\frac{\dot{f}}{f})\delta Q\Big)\partial^2\mathbb{Y}+2\sqrt{2}{a^2}f_{\phi}\frac{\dot{\mathbb{A}}}{a}\Big[3\delta\dot{Q}+3(H-\frac{\dot{f}}{f})\delta Q\Big]\nonumber\\
    &-4ff_{\phi}\frac{\dot{\mathbb{A}}}{a^2}\partial^2\mathbb{Y}\delta\phi+\frac{3}{a^2}\big(f_{\phi}^2+ff_{\phi\phi}\big)\mathbb{A}^2(\delta\phi)^2\\
     &+\Big[\Big(12H\dot{\zeta}-\frac{4}{a^2}\partial^2\zeta\Big)M_{\text{pl}}^2+6f^2\frac{\dot{\mathbb{A}}^2}{a^2}\zeta\Big]A+4M_{\text{pl}}^2\frac{1}{a^2}\dot{\zeta}\partial^2\beta\nonumber\\
     &-\Big(6\dot{\zeta}^2+36\dot{\zeta}\zeta+54h^2\zeta^2+\frac{2}{a^2}\zeta\partial^2\zeta\Big)M_{\text{pl}}^2+6\zeta\Big[\dot{\phi}-V_{\phi}\delta\phi+ff_{\phi}\frac{\dot{\mathbb{A}}^2}{a^2}\delta\phi\nonumber\\
     &+\sqrt{2}f\frac{\dot{\mathbb{A}}}{2a}\Big(\delta\dot{Q}+(H-\frac{\dot{f}}{f})\delta Q\Big)-f^2\frac{\dot{\mathbb{A}}}{3a^2}\partial^2\mathbb{Y}+3\zeta^2\Big(3\dot{\phi}^2+5f^2\frac{\dot{\mathbb{A}}^2}{a^2}\Big)\Big]\Bigg\},
\end{align}
where (A.1) is the parts which do not contain curvature perturbation $\zeta$ while (A.2) do. After imposing the comoving gauge (\ref{comovinggauge}), we can eliminate the inflaton perturbation $\delta \phi$. Then the equations of motion of non-dynamical degrees of freedom $\alpha$, $\beta$ and $\mathbb{Y}$ are
\begin{align}
    \alpha=\frac{\dot{\zeta}}{H},\ \ \ \ \ \ \ &\partial^2\mathbb{Y}=\dot{\mathbb{A}}\left(\zeta-\frac{\dot{\zeta}}{H}\right)+\frac{a}{\sqrt{2}f}\left[\delta\dot{Q}+\left(H-\frac{\dot{f}}{f}+4f_{\phi}\frac{\dot{\mathbb{A}}}{a}h\right)\delta Q\right],\nonumber\\
    \frac{4}{a^2}M_{\text{pl}}^2H\partial^2 \beta=&-4M_{\text{pl}}^2H^2\left(3-\epsilon_{\phi}-\frac{3}{2}\epsilon_A\right)\alpha-M_{\text{pl}}H^2\bigg[-12\sqrt{2}\frac{f_{\phi}}{f}M_{\text{pl}}\epsilon_Ah\nonumber\\
    &+2\sqrt{2}\sqrt{2\epsilon_{\phi}}(3+\eta_{\phi})h+3\sqrt{2\epsilon_A}\left(1-\frac{\dot{f}}{fH}\right)\bigg]\delta Q\nonumber\\
    &+M_{\text{pl}}H\sqrt{2\epsilon_A}\left(-\delta\dot{Q}+2H\eta_h\delta Q+\sqrt{2}\frac{f}{a}\partial^2\mathbb{Y}\right)\nonumber\\
    &+M_{\text{pl}}^2\left(12H\dot{\zeta}-\frac{4}{a^2}\partial^2\zeta\right)+6M_{\text{pl}}^2H^2\epsilon_A\zeta.
\end{align}
After inserting these solutions into the quadratic action and imposing a series of tedious integral by parts, the quadratic action can be written as (\ref{L2}), 
where the mass $m_{\sigma}$, $m_s$ and the coupling $\mathcal{A}$, $\mathcal{B}$ are 
\begin{align}
    &\ \ \ \ \ \ \ \ \mathcal{A}= 4+\mathcal{O}(\epsilon),\ \ \ \ \ \ \ \mathcal{B}= \frac{16h^2}{1+2h^2}+\mathcal{O}(\epsilon),\ \ \ \ \ \ \ m_{\sigma}^2=\frac{16h^2}{1+2h^2}H^2,\nonumber\\
    &m_s^2= \frac{8h^4-40h^2-4}{1+2h^2}H^2-\frac{12h^4-8h^2-2}{1+2h^2}\frac{f_{\phi\phi}}{f}M_{\text{pl}}^2H^2\epsilon_{\phi}+\frac{2h^2}{1+2h^2}V_{\phi\phi}+\mathcal{O}(\epsilon).
\end{align}
All coefficients except $m_{\sigma}$ have slow-roll corrections $\mathcal{O}(\epsilon)$. Here $\epsilon$ denotes all kinds of slow-roll parameter, i.e., $\epsilon$, $\epsilon_{\phi}$, $\epsilon_{A}$ and their higher derivative of time. We consider decoupling limit in this paper hence these corrections can be discarded. Also, we assume the slow-roll condition that $V_{\phi\phi}\ll H^2$ hence the third term in $m_s^2$ also can be ignored.

\section{Modified dispersion relation}
\label{app:EFT}
We are interested in the decoupling limit where the nontrivial solution $\zeta=\text{constant}$ is permitted hence we drop out the small slow-roll parameters. Then we have $\mathcal{A}=4$ and $\mathcal{B}=m_{\sigma}/H^2$. We regard $\dot{\sigma}$ and $h$ as constants. To find the dispersion relation we study the short-wavelength limit where the Hubble friction terms can be disregarded. We have noted that there is also a mass term for curvature fluctuation. The upper bound of the mass is $m_{\sigma}^2\lesssim 8H^2$, i.e., the mass of $\zeta$ is at most of the same order as the energy scale of inflation. In the short-wavelength limit the contribution of $m_{\sigma}$ term can also be ignored. Then the equations of motion read
\begin{align}\label{seom}
    \ddot{\zeta_c}+p^2\zeta_c=&\sqrt{2}hH\mathcal{A}\dot{\mathcal{F}},\nonumber\\
    \ddot{\mathcal{F}}+p^2\mathcal{F}+m_{s}^2\mathcal{F}=&-\sqrt{2}hH\mathcal{A}\dot{\zeta}_c,
\end{align}
where $p=k/a$ is the physical momentum and $\zeta_c\equiv M_{\text{pl}}\sqrt{2\epsilon}\zeta$ is the canonical variable. The schematic solutions of these equations can be written as \cite{Achucarro:2012yr}
\begin{align}
    &\zeta_c= \zeta_+e^{i\omega_+t}+\zeta_-e^{i\omega_-t},\nonumber\\
    &\mathcal{F}= \mathcal{F}_+e^{i\omega_+t}+\mathcal{F}_-e^{i\omega_-t},
\end{align}
where the two frequencies $\omega_{\pm}$ are given by 
\begin{align} \label{omega}
    \omega_{\pm}^2= \frac{m_s^2}{2c_s^2}+p^2\pm\frac{m_s^2}{2c_s^2}\sqrt{1+\frac{4\left(1-c_s^2\right)p^2}{m_s^2c_s^{-2}}},
\end{align}
where the speed of sound is given by
\begin{equation}
    \frac{1}{c_s^2}= 1+\frac{32H^2h^2}{m_s^2}.
\end{equation}
The speed of sound can be given by the dispersion relation of low-frequency modes at low-energy $4(1-c_s^2)p^2\ll m_s^2 c_s^{-2}$
\begin{align}
    \omega_-^2\simeq c_s^2p^2.
\end{align}
The modification of the dispersion relation of fluctuations is provided by the interaction with $\dot{\zeta}{\mathcal{F}}$. The heavy entropy fluctuation implies the hierarchy of the two frequencies $\omega_-\ll\omega_+$. Hence we have a cut-off scale on momentum
\begin{equation}\label{pUV}
    p_{\text{UV}}^2=\frac{|m_s^2 c_s^{-2}|}{4\left(1-c_s^2\right)}.
\end{equation}
In the low energy scales, the two scalar modes, which are corresponded to two frequencies, oscillate coherently and we integrate out the heavy mode to eliminate the high-frequency mode from the dynamics of the curvature fluctuation.

\section{Growing solution of the adiabatic mode}
\label{app:I}

The negative mass square of entropy fluctuation may lead to a imaginary low frequency $\omega_{-}$. Then the curvature fluctuation start to exponentially grow at some time $\tilde{N}_k$, which is determined by $\omega_{-}(\tilde{N}_k)=0$: $p^2(\tilde{N}_k)=-m_{s}^2$. If we neglect the slow rolling, the power spectrum is scale invariant on large scales and we can rescale it as $p/H=e^{-N}$. The growing solution of the curvature fluctuation can be written as \cite{Bjorkmo:2019qno}
\begin{align}\label{azeta}
    \zeta\propto \exp{\left[\int^{N}_{-\ln{(|m_s|/H)}}\ \frac{|\omega_{-}|}{H} dN\right]},
\end{align}
where $\omega_{-}$ is given by (\ref{omega}) 
\begin{equation}
    \frac{|\omega_{-}|}{H}=\sqrt{-\mu^2-\frac{1}{2}u+\frac{1}{2}\sqrt{u^2+128h^2\mu^2}}
\end{equation}
and $\mu\equiv p/H$ and $u\equiv m_s^2/H^2+32h^2$. It's useful to define the variable $y$ as $u^2y^2=u^2+128h^2\mu^2$ Then the integration in (\ref{azeta}) becomes
\begin{equation}
    \mathcal{I}(N)\equiv\frac{u}{8\sqrt{2}h}\int_{y_1}^{y_2}\sqrt{b\left(y-1\right)-\left(y^2-1\right)}\frac{ydy}{y^2-1}
\end{equation}
where the parameter $b\equiv 64h^2/u$, $c\equiv\lambda/u$ and the limits are
\begin{equation}
    y_1=\sqrt{1+\frac{128h^2\mu^2}{u^2}},\ \ \ \ \ \ \ y_2=\sqrt{1+\frac{128h^2|m_s|^2}{u^2H^2}}.
\end{equation}
The primitive function above can be calculated and given by
\begin{align}\label{Fy}
    F(y)=&\frac{u}{8\sqrt{2}h}\Bigg\{\sqrt{1-b+by-y^2}-\frac{b}{2}\arctan\left[\frac{b-2y}{2\sqrt{1-b+by-y^2}}\right]\nonumber\\
    &-\frac{\sqrt{2b}}{2}\arctan\left[\frac{2(1-y)+b(y-1-2)}{2\sqrt{2b}\sqrt{1-b+by-y^2}}\right],
\end{align}
For two $\arctan x$ in $F(y)$ at $y_2$, the first one is negative $-\pi/2$ while the second one is positive $\pi/2$ due to $b-2y_2<0$ and $2(1-y_2)+b(y_2)-1-2c>0$. On the other hand, the modes will be frozen-in when crossing the horizon hence we take the upper limit at $k|c_s|=aH$, i.e., 
\begin{equation}
    y_1=\sqrt{1+\frac{128h^2}{u^2|c_s|^2}}
\end{equation}
Summarizing the above computation, we finally have
\begin{align}
    \mathcal{I}=\frac{u}{8\sqrt{2}h}\Bigg\{&-\sqrt{1-b+by_1-y_1^2}+\frac{b}{2}\left[\frac{\pi}{2}+\arctan\left[\frac{b-2y_1}{2\sqrt{1-b+by_1-y_1^2}}\right]\right]\nonumber\\
    &-\frac{\sqrt{2b}}{2}\left[\frac{\pi}{2}-\arctan\left[\frac{2(1-y_1)+b(y_1-3)}{2\sqrt{2b}\sqrt{1-b+by_1-y_1^2}}\right]\right]\Bigg\}.
\end{align}
We can simple take the super-horizon limit $\mu\to 0$ and in the slow-roll limit,
\begin{equation}
    u^2\simeq \frac{8h^2(1+6h^2)}{1+2h^2},  \ \ \ \ \ \ \ b\simeq \frac{8(1+2h^2)}{1+6h^2},\ \ \ \ \ \ \ y_1\simeq 1.
\end{equation}
Then the function $\mathcal{I}$ can be reduced to a simply form
\begin{equation}
    \mathcal{I}_{\text{approx}}= \sqrt{2}\pi\left(2-\sqrt{\frac{1+6h^2}{1+2h^2}}\right)h.
\end{equation}
For large $h\gg 1$ we found $\mathcal{I}_{\text{approx}}\simeq 1.186h$, which is very closed to the numerical result $1.193h$ we have found in \cite{Chen:2022ccf}. We found that the approximation result is good enough for $h\gtrsim 5$. 

\section{Bogoliubov coefficients}
\label{app:Bogoliubov}

For the sharp case, we use the top-hat profile of the $h(N)$ for analytical computations,
\begin{equation}
    h(N)=h_0\left[\theta(t-t_1)-\theta(t-t_2)\right].
\end{equation}
Before $t_1$,  where we denote as region-I, the curvature and entropy modes are decoupled. Hence the solutions are simply the mode function with Bunch-Davies initial conditions, as the same as the single-field case,
\begin{align}
    \hat{\zeta}_c^{\text{I}}(t,\boldsymbol{k})=&u_k(t)\hat{a}_{\zeta}(\boldsymbol{k})+\text{h.c.}(-\boldsymbol{k}),\nonumber\\
    \hat{\mathcal{F}}^{\text{I}}(t,\boldsymbol{k})=&u_k(t)\hat{a}_{\mathcal{F}}(\boldsymbol{k})+\text{h.c.}(-\boldsymbol{k}),
\end{align}
where $u_k$ is given by (\ref{BD}). 

After $t>t_1$, which is denoted as region-II, these two scalar fields are coupled to each other hence their modes are also mixed with each other. The governing equations of motion are given by (\ref{fulleom}). We discuss two different non-trivial momentum regimes: near PBHs peak $k\sim k_f$ and large scales $k\ll k_f$, respectively. In order to preserve the continuity of the commutation relations, we should impose the boundary conditions at $t=t_1$ as
\begin{align}
    \hat{\zeta}_c^{\text{I}}(t_1^-)=&\hat{\zeta}_c^{\text{II}}(t_1^+),\ \ \ \ \ \ \ \dot{\hat{\zeta}}_c^{\text{I}}(t_1^-)=D_t\hat{\zeta}_c^{\text{II}}(t_1^+),\nonumber\\
    \hat{\mathcal{F}}^{\text{I}}(t_1^-)=&\hat{\mathcal{F}}^{\text{II}}(t_1^+),\ \ \ \ \ \ \ \dot{\hat{\mathcal{F}}}^{\text{I}}(t_1^-)=\dot{\hat{\mathcal{F}}}^{\text{II}}(t_1^+),
\end{align}
where $D_t\hat{\zeta}_c\equiv \dot{\hat{\zeta}}_c-\sqrt{2}h\mathcal{A}H\hat{\mathcal{F}}$.

During region-III $t>t_2$, where the curvature and entropy modes are decoupled again, the mode functions are again the ones with Bunch-David vacuum but mixing these two scalar modes,
\begin{align}
    \hat{\zeta}_c^{\text{III}}(t,\boldsymbol{k})=&\Big[E_{\zeta}u_k(t)+F_{\zeta}u_k^*(t)\Big]\hat{a}_{\zeta}(\boldsymbol{k})+\Big[G_{\zeta}u_k(t)+H_{\zeta}u_k^*(t)\Big]\hat{a}_{\mathcal{F}}(\boldsymbol{k})\nonumber\\
    &+\text{h.c.}(-\boldsymbol{k}),\nonumber\\
    \hat{\mathcal{F}}^{\text{III}}(t,\boldsymbol{k})=&\Big[E_{\mathcal{F}}u_k(t)+F_{\mathcal{F}}u_k^*(t)\Big]\hat{a}_{\zeta}(\boldsymbol{k})+\Big[G_{\mathcal{F}}u_k(t)+H_{\mathcal{F}}u_k^*(t)\Big]\hat{a}_{\mathcal{F}}(\boldsymbol{k})\nonumber\\
    &+\text{h.c.}(-\boldsymbol{k}).
\end{align}
We then also impose the boundary conditions at $t=t_2$ as
\begin{align}
    \hat{\zeta}_c^{\text{II}}(t_2^-)=&\hat{\zeta}_c^{\text{III}}(t_2^+),\ \ \ \ \ \ \ D_t\hat{\zeta}_c^{\text{II}}(t_2^-)=\dot{\hat{\zeta}}_c^{\text{III}}(t_2^+),\nonumber\\
    \hat{\mathcal{F}}^{\text{II}}(t_2^-)=&\hat{\mathcal{F}}^{\text{III}}(t_2^+),\ \ \ \ \ \ \ \dot{\hat{\mathcal{F}}}^{\text{II}}(t_2^-)=\dot{\hat{\mathcal{F}}}^{\text{III}}(t_2^+).
\end{align}
We have eight coefficients $E_i$, $F_i$, $G_i$ and $H_i$, where $i=\zeta,\mathcal{F}$ and can be determined by the eight boundary conditions at $t_2$. 

For modes with $k\gtrsim k_f$, the coefficients of curvature perturbations are solved to
\begin{align}\label{EFGHfull}
    E_{\zeta}=&\frac{e^{i2\kappa\sinh\left(\frac{\delta N}{2}\right)}}{4\kappa\left(S_+^2-S_-^2\right)}\sum_{\pm}\pm\frac{S_{\mp}^2-\kappa^2}{S_{\pm}}\Bigg\{e^{iS_{\pm}\delta N}\left[1+\left(S_{\pm}-\kappa\right)^2\right]-e^{-iS_{\pm}\delta N}\left[1+\left(S_{\pm}+\kappa\right)^2\right]\Bigg\},\nonumber\\
    F_{\zeta}=&\frac{e^{i2\kappa\cosh\left(\frac{\delta N}{2}\right)}}{2\kappa\left(S_+^2-S_-^2\right)}\sum_{\pm}\pm\frac{i \sin{\left(S_{\pm}\delta N\right)}}{S_{\pm}}\left(S_{\mp}^2-\kappa^2\right)\left[S_{\pm}^2+\left(1-\kappa^2-i2\kappa\right)\right],
    \nonumber\\
    G_{\zeta}=&\frac{i\sqrt{2}h\mathcal{A}e^{i2\kappa\sinh\left(\frac{\delta N}{2}\right)}}{4\kappa\left(S_+^2-S_-^2\right)}\sum_{\pm}\pm\frac{1}{S_{\pm}}\Bigg\{e^{iS_{\pm}\delta N}\left[\kappa^2-S_{\pm}(i+\kappa)\right]\left(i-\kappa+S_{\pm}\right)\nonumber\\
     &\ \ \ \ \ \ \ \ \ \ \ \ \ \ \ \ \ \ \ \ \ \ \ \ \ \ \ \ \ \ \ \ \ \ \ \ \ \ \ \ -e^{-iS_{\pm}\delta N}\left[\kappa^2+S_{\pm}\left(i+\kappa\right)\right]\left(i-\kappa-S_{\pm}\right)\Bigg\},\nonumber\\
     H_{\zeta}=&\frac{i\sqrt{2}h\mathcal{A}e^{i2\kappa\cosh\left(\frac{\delta N}{2}\right)}}{4\kappa\left(S_+^2-S_-^2\right)}\sum_{\pm}\pm\frac{1}{S_{\pm}}\Bigg\{e^{iS_{\pm}\delta N}\left[\kappa^2-S_{\pm}(i+\kappa)\right]\left(i+\kappa+S_{\pm}\right)\nonumber\\
     &\ \ \ \ \ \ \ \ \ \ \ \ \ \ \ \ \ \ \ \ \ \ \ \ \ \ \ \ \ \ \ \ \ \ \ \ \ \ \ \ -e^{-iS_{\pm}\delta N}\left[\kappa^2+S_{\pm}\left(i+\kappa\right)\right]\left(i+\kappa-S_{\pm}\right)\Bigg\},
\end{align}
where $\kappa\equiv k/k_f$, $\delta N\equiv H(t_2-t_1)$ and $S_{\pm}\equiv \omega_{\pm}/H$.

For modes with $k\ll k_f$, the coefficients can be also solved. The final results depends on $h$, $\delta N$ and $x_2=-k\tau_2$ and are extremely tedious. This is because the matching on boundaries is not exact. However we can simple the results under some approximations. Firstly, we are interested in the large $h$ case and we can expand the results as small $1/h$. Secondly, we are considering the super-horizon limit hence $x_2\ll 1$. Then we can also expand the results as small $x_2$.  The dominated terms up to first term in small $h$ and first two terms in small $x_2$ in these two expansions are
\begin{align}
    E_{\zeta}=&\frac{1}{27}\left[32\cosh{\left(3\delta N\right)}-5\right]+i\frac{4\left(1-e^{-3\delta N}\right)}{9x_2^3}
    +i\frac{e^{-\delta N/2}}{3x_2}\bigg[\sinh\left(\frac{\delta N}{2}\right)\nonumber\\
    &+4\sinh\left(\frac{5\delta N}{2}\right)\bigg]+\cdots,\nonumber\\
    F_{\zeta}=&\frac{32}{27}\sinh\left(3\delta N\right)+i\frac{4\left(1-e^{-3\delta N}\right)}{9x_2^3}+i\frac{e^{-\delta N/2}}{3x_2}\left[\sinh\left(\frac{\delta N}{2}\right)+4\sinh\left(\frac{5\delta N}{2}\right)\right]+\cdots,\nonumber\\
    G_{\zeta}=&\frac{e^{-3\delta N/2}}{3\sqrt{3}}\left(2e^{3\delta N}+1-i\frac{6}{x_2^3}-i e^{2\delta N}\frac{3}{x_2}\right)\sin{\left(2\sqrt{6}h \delta N\right)}+\cdots,\nonumber\\
    H_{\zeta}=&\frac{e^{-3\delta N/2}}{3\sqrt{3}}\left(2e^{3\delta N}-1-i\frac{6}{x_2^3}-i e^{2\delta N}\frac{3}{x_2}\right)\sin{\left(2\sqrt{6}h \delta N\right)}+\cdots.
\end{align}
We find that although the results are complex and the imaginary parts are the functions of $x_2$, in $(E_{\zeta}-F_{\zeta})$ and $(G_{\zeta}-H_{\zeta})$, the imaginary parts will be killed to each other. Because perturbations $\zeta$ becomes classical and is a constant after horizon-crossing.


\begin{thebibliography}{99}

\bibitem{Smoot:1992}
G. F. Smoot et al.,
``Structure in the COBE Differential Microwave Radiometer First-Year Maps'',
Astrophys. J. \textbf{396}, L1 (1992)

\bibitem{WMAP:2003elm}
D.~N.~Spergel \textit{et al.} [WMAP],
``First year Wilkinson Microwave Anisotropy Probe (WMAP) observations: Determination of cosmological parameters,''
Astrophys. J. Suppl. \textbf{148}, 175-194 (2003)
doi:10.1086/377226
[arXiv:astro-ph/0302209 [astro-ph]].

\bibitem{Planck:2018vyg}
N.~Aghanim \textit{et al.} [Planck],
``Planck 2018 results. VI. Cosmological parameters,''
Astron. Astrophys. \textbf{641}, A6 (2020)
[erratum: Astron. Astrophys. \textbf{652}, C4 (2021)]
doi:10.1051/0004-6361/201833910
[arXiv:1807.06209 [astro-ph.CO]].

\bibitem{Copeland:1994vg}
E.~J.~Copeland, A.~R.~Liddle, D.~H.~Lyth, E.~D.~Stewart and D.~Wands,
``False vacuum inflation with Einstein gravity,''
Phys. Rev. D \textbf{49}, 6410-6433 (1994)
doi:10.1103/PhysRevD.49.6410
[arXiv:astro-ph/9401011 [astro-ph]].

\bibitem{Baumann:2014nda}
D.~Baumann and L.~McAllister,
``Inflation and String Theory,''
Cambridge University Press, 2015,
ISBN 978-1-107-08969-3, 978-1-316-23718-2
doi:10.1017/CBO9781316105733
[arXiv:1404.2601 [hep-th]].

\bibitem{Lyth:1996im}
D.~H.~Lyth,
``What would we learn by detecting a gravitational wave signal in the cosmic microwave background anisotropy?,''
Phys. Rev. Lett. \textbf{78}, 1861-1863 (1997)
doi:10.1103/PhysRevLett.78.1861
[arXiv:hep-ph/9606387 [hep-ph]].

\bibitem{Baumann:2011ws}
D.~Baumann and D.~Green,
``A Field Range Bound for General Single-Field Inflation,''
JCAP \textbf{05}, 017 (2012)
doi:10.1088/1475-7516/2012/05/017
[arXiv:1111.3040 [hep-th]].

\bibitem{Brown:2017osf}
A.~R.~Brown,
``Hyperbolic Inflation,''
Phys. Rev. Lett. \textbf{121}, no.25, 251601 (2018)
doi:10.1103/PhysRevLett.121.251601
[arXiv:1705.03023 [hep-th]].

\bibitem{Mizuno:2017idt}
S.~Mizuno and S.~Mukohyama,
``Primordial perturbations from inflation with a hyperbolic field-space,''
Phys. Rev. D \textbf{96}, no.10, 103533 (2017)
doi:10.1103/PhysRevD.96.103533
[arXiv:1707.05125 [hep-th]].

\bibitem{Christodoulidis:2018qdw}
P.~Christodoulidis, D.~Roest and E.~I.~Sfakianakis,
``Angular inflation in multi-field $\alpha$-attractors,''
JCAP \textbf{11}, 002 (2019)
doi:10.1088/1475-7516/2019/11/002
[arXiv:1803.09841 [hep-th]].

\bibitem{Renaux-Petel:2015mga}
S.~Renaux-Petel and K.~Turzy\'nski,
``Geometrical Destabilization of Inflation,''
Phys. Rev. Lett. \textbf{117}, no.14, 141301 (2016)
doi:10.1103/PhysRevLett.117.141301
[arXiv:1510.01281 [astro-ph.CO]].

\bibitem{Garcia-Saenz:2018ifx}
S.~Garcia-Saenz, S.~Renaux-Petel and J.~Ronayne,
``Primordial fluctuations and non-Gaussianities in sidetracked inflation,''
JCAP \textbf{07}, 057 (2018)
doi:10.1088/1475-7516/2018/07/057
[arXiv:1804.11279 [astro-ph.CO]].

\bibitem{Bjorkmo:2019fls}
T.~Bjorkmo,
``Rapid-Turn Inflationary Attractors,''
Phys. Rev. Lett. \textbf{122}, no.25, 251301 (2019)
doi:10.1103/PhysRevLett.122.251301
[arXiv:1902.10529 [hep-th]].

\bibitem{Fumagalli:2019noh}
J.~Fumagalli, S.~Garcia-Saenz, L.~Pinol, S.~Renaux-Petel and J.~Ronayne,
``Hyper-Non-Gaussianities in Inflation with Strongly Nongeodesic Motion,''
Phys. Rev. Lett. \textbf{123}, no.20, 201302 (2019)
doi:10.1103/PhysRevLett.123.201302
[arXiv:1902.03221 [hep-th]].

\bibitem{Planck:2018jri}
Y.~Akrami \textit{et al.} [Planck],
``Planck 2018 results. X. Constraints on inflation,''
Astron. Astrophys. \textbf{641}, A10 (2020)
doi:10.1051/0004-6361/201833887
[arXiv:1807.06211 [astro-ph.CO]].

\bibitem{Tavecchio:2010}
F. Tavecchio, et al.,
``The intergalactic magnetic field constrained by Fermi/Large Area Telescope observations of the TeV blazar 1ES 0229+ 200,''
Mon. Not. R. Astron. Soc. \textbf{406}, L70 (2010).

\bibitem{Neronov:2010}
A. Neronov and I. Vovk,
``Evidence for strong extragalactic magnetic fields from Fermi observations of TeV blazars,''
Science \textbf{328}, 73 (2010).

\bibitem{Tavecchio:2011}
F. Tavecchio, G. Ghisellini, G. Bonnoli and L. Foschini,
``Extreme TeV blazars and the intergalactic magnetic field,''
Mon. Not. R. Astron. Soc. \textbf{414}, 3566 (2011).

\bibitem{Subramanian:2015lua}
K.~Subramanian,
``The origin, evolution and signatures of primordial magnetic fields,''
Rept. Prog. Phys. \textbf{79}, no.7, 076901 (2016)
doi:10.1088/0034-4885/79/7/076901
[arXiv:1504.02311 [astro-ph.CO]].

\bibitem{Chen:2022ccf}
C.~B.~Chen and J.~Soda,
``Geometric structure of multi-form-field isotropic inflation and primordial fluctuations,''
JCAP \textbf{05}, no.05, 029 (2022)
doi:10.1088/1475-7516/2022/05/029
[arXiv:2201.03160 [hep-th]].

\bibitem{Watanabe:2009ct}
M.~a.~Watanabe, S.~Kanno and J.~Soda,
``Inflationary Universe with Anisotropic Hair,''
Phys. Rev. Lett. \textbf{102}, 191302 (2009)
doi:10.1103/PhysRevLett.102.191302
[arXiv:0902.2833 [hep-th]].

\bibitem{Watanabe:2010fh}
M.~a.~Watanabe, S.~Kanno and J.~Soda,
``The Nature of Primordial Fluctuations from Anisotropic Inflation,''
Prog. Theor. Phys. \textbf{123}, 1041-1068 (2010)
doi:10.1143/PTP.123.1041
[arXiv:1003.0056 [astro-ph.CO]].

\bibitem{Kanno:2010nr}
S.~Kanno, J.~Soda and M.~a.~Watanabe,
``Anisotropic Power-law Inflation,''
JCAP \textbf{12}, 024 (2010)
doi:10.1088/1475-7516/2010/12/024
[arXiv:1010.5307 [hep-th]].

\bibitem{Yamamoto:2012tq}
K.~Yamamoto, M.~a.~Watanabe and J.~Soda,
``Inflation with Multi-Vector-Hair: The Fate of Anisotropy,''
Class. Quant. Grav. \textbf{29}, 145008 (2012)
doi:10.1088/0264-9381/29/14/145008
[arXiv:1201.5309 [hep-th]].

\bibitem{Maleknejad:2012fw}
A.~Maleknejad, M.~M.~Sheikh-Jabbari and J.~Soda,
``Gauge Fields and Inflation,''
Phys. Rept. \textbf{528}, 161-261 (2013)
doi:10.1016/j.physrep.2013.03.003
[arXiv:1212.2921 [hep-th]].

\bibitem{Bartolo:2012sd}
N.~Bartolo, S.~Matarrese, M.~Peloso and A.~Ricciardone,
``Anisotropic power spectrum and bispectrum in the $f(\phi)F^2$ mechanism,''
Phys. Rev. D \textbf{87}, no.2, 023504 (2013)
doi:10.1103/PhysRevD.87.023504
[arXiv:1210.3257 [astro-ph.CO]].

\bibitem{Naruko:2014bxa}
A.~Naruko, E.~Komatsu and M.~Yamaguchi,
``Anisotropic inflation reexamined: upper bound on broken rotational invariance during inflation,''
JCAP \textbf{04}, 045 (2015)
doi:10.1088/1475-7516/2015/04/045
[arXiv:1411.5489 [astro-ph.CO]].

\bibitem{Fujita:2017lfu}
T.~Fujita and I.~Obata,
``Does anisotropic inflation produce a small statistical anisotropy?,''
JCAP \textbf{01}, 049 (2018)
doi:10.1088/1475-7516/2018/01/049
[arXiv:1711.11539 [astro-ph.CO]].

\bibitem{Gorji:2020vnh}
M.~A.~Gorji, S.~A.~Hosseini Mansoori and H.~Firouzjahi,
``Inflation with multiple vector fields and non-Gaussianities,''
JCAP \textbf{11}, 041 (2020)
doi:10.1088/1475-7516/2020/11/041
[arXiv:2008.08195 [astro-ph.CO]].

\bibitem{Yamamoto:2012sq}
K.~Yamamoto,
``Primordial Fluctuations from Inflation with a Triad of Background Gauge Fields,''
Phys. Rev. D \textbf{85}, 123504 (2012)
doi:10.1103/PhysRevD.85.123504
[arXiv:1203.1071 [astro-ph.CO]].

\bibitem{Funakoshi:2012ym}
H.~Funakoshi and K.~Yamamoto,
``Primordial bispectrum from inflation with background gauge fields,''
Class. Quant. Grav. \textbf{30}, 135002 (2013)
doi:10.1088/0264-9381/30/13/135002
[arXiv:1212.2615 [astro-ph.CO]].

\bibitem{Arnowitt:1962hi}
R.~L.~Arnowitt, S.~Deser and C.~W.~Misner,
``The Dynamics of general relativity,''
Gen. Rel. Grav. \textbf{40}, 1997-2027 (2008)
doi:10.1007/s10714-008-0661-1
[arXiv:gr-qc/0405109 [gr-qc]].

\bibitem{Bento:1992wy}
M.~C.~Bento, O.~Bertolami, P.~V.~Moniz, J.~M.~Mourao and P.~M.~Sa,
``On the cosmology of massive vector fields with SO(3) global symmetry,''
Class. Quant. Grav. \textbf{10}, 285-298 (1993)
doi:10.1088/0264-9381/10/2/010
[arXiv:gr-qc/9302034 [gr-qc]].

\bibitem{Golovnev:2008cf}
A.~Golovnev, V.~Mukhanov and V.~Vanchurin,
``Vector Inflation,''
JCAP \textbf{06}, 009 (2008)
doi:10.1088/1475-7516/2008/06/009
[arXiv:0802.2068 [astro-ph]].

\bibitem{Murata:2011wv}
K.~Murata and J.~Soda,
``Anisotropic Inflation with Non-Abelian Gauge Kinetic Function,''
JCAP \textbf{06}, 037 (2011)
doi:10.1088/1475-7516/2011/06/037
[arXiv:1103.6164 [hep-th]].

\bibitem{Maleknejad:2011sq}
A.~Maleknejad and M.~M.~Sheikh-Jabbari,
``Non-Abelian Gauge Field Inflation,''
Phys. Rev. D \textbf{84}, 043515 (2011)
doi:10.1103/PhysRevD.84.043515
[arXiv:1102.1932 [hep-ph]].

\bibitem{Maleknejad:2011jw}
A.~Maleknejad and M.~M.~Sheikh-Jabbari,
``Gauge-flation: Inflation From Non-Abelian Gauge Fields,''
Phys. Lett. B \textbf{723}, 224-228 (2013)
doi:10.1016/j.physletb.2013.05.001
[arXiv:1102.1513 [hep-ph]].

\bibitem{Achucarro:2012sm}
A.~Achucarro, J.~O.~Gong, S.~Hardeman, G.~A.~Palma and S.~P.~Patil,
``Effective theories of single field inflation when heavy fields matter,''
JHEP \textbf{05}, 066 (2012)
doi:10.1007/JHEP05(2012)066
[arXiv:1201.6342 [hep-th]].

\bibitem{Garcia-Saenz:2019njm}
S.~Garcia-Saenz, L.~Pinol and S.~Renaux-Petel,
``Revisiting non-Gaussianity in multifield inflation with curved field space,''
JHEP \textbf{01}, 073 (2020)
doi:10.1007/JHEP01(2020)073
[arXiv:1907.10403 [hep-th]].

\bibitem{Firouzjahi:2018wlp}
H.~Firouzjahi, M.~A.~Gorji, S.~A.~Hosseini Mansoori, A.~Karami and T.~Rostami,
``Charged Vector Inflation,''
Phys. Rev. D \textbf{100}, no.4, 043530 (2019)
doi:10.1103/PhysRevD.100.043530
[arXiv:1812.07464 [hep-th]].

\bibitem{Achucarro:2016fby}
A.~Ach\'ucarro, V.~Atal, C.~Germani and G.~A.~Palma,
``Cumulative effects in inflation with ultra-light entropy modes,''
JCAP \textbf{02}, 013 (2017)
doi:10.1088/1475-7516/2017/02/013
[arXiv:1607.08609 [astro-ph.CO]].

\bibitem{Cheung:2007st}
C.~Cheung, P.~Creminelli, A.~L.~Fitzpatrick, J.~Kaplan and L.~Senatore,
``The Effective Field Theory of Inflation,''
JHEP \textbf{03}, 014 (2008)
doi:10.1088/1126-6708/2008/03/014
[arXiv:0709.0293 [hep-th]].

\bibitem{Gwyn:2012mw}
R.~Gwyn, G.~A.~Palma, M.~Sakellariadou and S.~Sypsas,
``Effective field theory of weakly coupled inflationary models,''
JCAP \textbf{04}, 004 (2013)
doi:10.1088/1475-7516/2013/04/004
[arXiv:1210.3020 [hep-th]].

\bibitem{Baumann:2011su}
D.~Baumann and D.~Green,
``Equilateral Non-Gaussianity and New Physics on the Horizon,''
JCAP \textbf{09}, 014 (2011)
doi:10.1088/1475-7516/2011/09/014
[arXiv:1102.5343 [hep-th]].

\bibitem{Achucarro:2012yr}
A.~Achucarro, V.~Atal, S.~Cespedes, J.~O.~Gong, G.~A.~Palma and S.~P.~Patil,
``Heavy fields, reduced speeds of sound and decoupling during inflation,''
Phys. Rev. D \textbf{86}, 121301 (2012)
doi:10.1103/PhysRevD.86.121301
[arXiv:1205.0710 [hep-th]].

\bibitem{Garcia-Saenz:2018vqf}
S.~Garcia-Saenz and S.~Renaux-Petel,
``Flattened non-Gaussianities from the effective field theory of inflation with imaginary speed of sound,''
JCAP \textbf{11}, 005 (2018)
doi:10.1088/1475-7516/2018/11/005
[arXiv:1805.12563 [hep-th]].


\bibitem{Garcia-Bellido:1995hsq}
J.~Garcia-Bellido and D.~Wands,
``Metric perturbations in two field inflation,''
Phys. Rev. D \textbf{53}, 5437-5445 (1996)
doi:10.1103/PhysRevD.53.5437
[arXiv:astro-ph/9511029 [astro-ph]].

\bibitem{Wands:2000dp}
D.~Wands, K.~A.~Malik, D.~H.~Lyth and A.~R.~Liddle,
``A New approach to the evolution of cosmological perturbations on large scales,''
Phys. Rev. D \textbf{62}, 043527 (2000)
doi:10.1103/PhysRevD.62.043527
[arXiv:astro-ph/0003278 [astro-ph]].

\bibitem{Finelli:2000ya}
F.~Finelli and R.~H.~Brandenberger,
``Parametric amplification of metric fluctuations during reheating in two field models,''
Phys. Rev. D \textbf{62}, 083502 (2000)
doi:10.1103/PhysRevD.62.083502
[arXiv:hep-ph/0003172 [hep-ph]].

\bibitem{Chen:2021nkf}
C.~B.~Chen and J.~Soda,
``Anisotropic hyperbolic inflation,''
JCAP \textbf{09}, 026 (2021)
doi:10.1088/1475-7516/2021/09/026
[arXiv:2106.04813 [hep-th]].

\bibitem{Zeldovich:1967lct}
Y.~B.~Zel'dovich and I.~D.~Novikov,
``The Hypothesis of Cores Retarded during Expansion and the Hot Cosmological Model,''
Soviet Astron. AJ (Engl. Transl. ), \textbf{10}, 602 (1967)

\bibitem{Hawking:1971ei}
S.~Hawking,
``Gravitationally collapsed objects of very low mass,''
Mon. Not. Roy. Astron. Soc. \textbf{152}, 75 (1971)
doi:10.1093/mnras/152.1.75

\bibitem{Carr:1974nx}
B.~J.~Carr and S.~W.~Hawking,
``Black holes in the early Universe,''
Mon. Not. Roy. Astron. Soc. \textbf{168}, 399-415 (1974)
doi:10.1093/mnras/168.2.399

\bibitem{Carr:1975qj}
B.~J.~Carr,
``The Primordial black hole mass spectrum,''
Astrophys. J. \textbf{201}, 1-19 (1975)
doi:10.1086/153853

\bibitem{Chapline:1975ojl}
G.~F.~Chapline,
``Cosmological effects of primordial black holes,''
Nature \textbf{253}, no.5489, 251-252 (1975)
doi:10.1038/253251a0

\bibitem{Carr:1984}
B.~J.~Carr and M.~J.~Rees, 
``Can pregalactic objects generate galaxies?,'' Mon. Not. Roy. Astron. Soc. \textbf{206}, 801–818 (1984)
doi: 10.1093/mnras/206.4.801 

\bibitem{Pi:2017gih}
S.~Pi, Y.~l.~Zhang, Q.~G.~Huang and M.~Sasaki,
``Scalaron from $R^2$-gravity as a heavy field,''
JCAP \textbf{05}, 042 (2018)
doi:10.1088/1475-7516/2018/05/042
[arXiv:1712.09896 [astro-ph.CO]].

\bibitem{Palma:2020ejf}
G.~A.~Palma, S.~Sypsas and C.~Zenteno,
``Seeding primordial black holes in multifield inflation,''
Phys. Rev. Lett. \textbf{125}, no.12, 121301 (2020)
doi:10.1103/PhysRevLett.125.121301
[arXiv:2004.06106 [astro-ph.CO]].

\bibitem{Fumagalli:2020adf}
J.~Fumagalli, S.~Renaux-Petel, J.~W.~Ronayne and L.~T.~Witkowski,
``Turning in the landscape: A new mechanism for generating primordial black holes,''
Phys. Lett. B \textbf{841}, 137921 (2023)
doi:10.1016/j.physletb.2023.137921
[arXiv:2004.08369 [hep-th]].

\bibitem{Bjorkmo:2019qno}
T.~Bjorkmo, R.~Z.~Ferreira and M.~C.~D.~Marsh,
``Mild Non-Gaussianities under Perturbative Control from Rapid-Turn Inflation Models,''
JCAP \textbf{12}, 036 (2019)
doi:10.1088/1475-7516/2019/12/036
[arXiv:1908.11316 [hep-th]].

\bibitem{Christodoulidis:2023eiw}
P.~Christodoulidis and J.~O.~Gong,
``Enhanced power spectra from multi-field inflation,''
[arXiv:2311.04090 [hep-th]].

\bibitem{Carr:2009jm}
B.~J.~Carr, K.~Kohri, Y.~Sendouda and J.~Yokoyama,
``New cosmological constraints on primordial black holes,''
Phys. Rev. D \textbf{81}, 104019 (2010)
doi:10.1103/PhysRevD.81.104019
[arXiv:0912.5297 [astro-ph.CO]].

\bibitem{Press:1973iz}
W.~H.~Press and P.~Schechter,
``Formation of galaxies and clusters of galaxies by selfsimilar gravitational condensation,''
Astrophys. J. \textbf{187}, 425-438 (1974)
doi:10.1086/152650

\bibitem{Harada:2013epa}
T.~Harada, C.~M.~Yoo and K.~Kohri,
``Threshold of primordial black hole formation,''
Phys. Rev. D \textbf{88}, no.8, 084051 (2013)
[erratum: Phys. Rev. D \textbf{89}, no.2, 029903 (2014)]
doi:10.1103/PhysRevD.88.084051
[arXiv:1309.4201 [astro-ph.CO]].

\bibitem{Young:2014ana}
S.~Young, C.~T.~Byrnes and M.~Sasaki,
``Calculating the mass fraction of primordial black holes,''
JCAP \textbf{07}, 045 (2014)
doi:10.1088/1475-7516/2014/07/045
[arXiv:1405.7023 [gr-qc]].

\bibitem{Sasaki:2018dmp}
M.~Sasaki, T.~Suyama, T.~Tanaka and S.~Yokoyama,
``Primordial black holes\textemdash{}perspectives in gravitational wave astronomy,''
Class. Quant. Grav. \textbf{35}, no.6, 063001 (2018)
doi:10.1088/1361-6382/aaa7b4
[arXiv:1801.05235 [astro-ph.CO]].

\bibitem{Niikura:2017zjd}
H.~Niikura, M.~Takada, N.~Yasuda, R.~H.~Lupton, T.~Sumi, S.~More, T.~Kurita, S.~Sugiyama, A.~More and M.~Oguri, \textit{et al.}
``Microlensing constraints on primordial black holes with Subaru/HSC Andromeda observations,''
Nature Astron. \textbf{3}, no.6, 524-534 (2019)
doi:10.1038/s41550-019-0723-1
[arXiv:1701.02151 [astro-ph.CO]].

\bibitem{Ando:2018qdb}
K.~Ando, K.~Inomata and M.~Kawasaki,
``Primordial black holes and uncertainties in the choice of the window function,''
Phys. Rev. D \textbf{97}, no.10, 103528 (2018)
doi:10.1103/PhysRevD.97.103528
[arXiv:1802.06393 [astro-ph.CO]].

\bibitem{Young:2019osy}
S.~Young,
``The primordial black hole formation criterion re-examined: Parametrisation, timing and the choice of window function,''
Int. J. Mod. Phys. D \textbf{29}, no.02, 2030002 (2019)
doi:10.1142/S0218271820300025
[arXiv:1905.01230 [astro-ph.CO]].

\bibitem{NANOGrav:2023gor}
G.~Agazie \textit{et al.} [NANOGrav],
``The NANOGrav 15 yr Data Set: Evidence for a Gravitational-wave Background,''
Astrophys. J. Lett. \textbf{951}, no.1, L8 (2023)
doi:10.3847/2041-8213/acdac6
[arXiv:2306.16213 [astro-ph.HE]].

\bibitem{NANOGrav:2023hvm}
A.~Afzal \textit{et al.} [NANOGrav],
``The NANOGrav 15 yr Data Set: Search for Signals from New Physics,''
Astrophys. J. Lett. \textbf{951}, no.1, L11 (2023)
doi:10.3847/2041-8213/acdc91
[arXiv:2306.16219 [astro-ph.HE]].

\bibitem{Kristiano:2022maq}
J.~Kristiano and J.~Yokoyama,
``Ruling Out Primordial Black Hole Formation From Single-Field Inflation,''
[arXiv:2211.03395 [hep-th]].

\bibitem{Riotto:2023hoz}
A.~Riotto,
``The Primordial Black Hole Formation from Single-Field Inflation is Not Ruled Out,''
[arXiv:2301.00599 [astro-ph.CO]].

\bibitem{Kristiano:2023scm}
J.~Kristiano and J.~Yokoyama,
``Response to criticism on ''Ruling Out Primordial Black Hole Formation From Single-Field Inflation'': A note on bispectrum and one-loop correction in single-field inflation with primordial black hole formation,''
[arXiv:2303.00341 [hep-th]].

\bibitem{Riotto:2023gpm}
A.~Riotto,
``The Primordial Black Hole Formation from Single-Field Inflation is Still Not Ruled Out,''
[arXiv:2303.01727 [astro-ph.CO]].

\bibitem{Firouzjahi:2023ahg}
H.~Firouzjahi and A.~Riotto,
``Primordial Black Holes and Loops in Single-Field Inflation,''
[arXiv:2304.07801 [astro-ph.CO]].

\bibitem{Choudhury:2023vuj}
S.~Choudhury, M.~R.~Gangopadhyay and M.~Sami,
``No-go for the formation of heavy mass Primordial Black Holes in Single Field Inflation,''
[arXiv:2301.10000 [astro-ph.CO]].

\bibitem{Choudhury:2023jlt}
S.~Choudhury, S.~Panda and M.~Sami,
``PBH formation in EFT of single field inflation with sharp transition,''
Phys. Lett. B \textbf{845}, 138123 (2023)
doi:10.1016/j.physletb.2023.138123
[arXiv:2302.05655 [astro-ph.CO]].

\bibitem{Choudhury:2023rks}
S.~Choudhury, S.~Panda and M.~Sami,
``Quantum loop effects on the power spectrum and constraints on primordial black holes,''
JCAP \textbf{11}, 066 (2023)
doi:10.1088/1475-7516/2023/11/066
[arXiv:2303.06066 [astro-ph.CO]].



\end{thebibliography}
\end{document}